\documentclass[twoside,twocolumn,9pt]{article}
\usepackage{extsizes}
\usepackage[super,sort&compress,comma]{natbib} 
\usepackage[version=3]{mhchem}
\usepackage[left=1.5cm, right=1.5cm, top=1.785cm, bottom=2.0cm]{geometry}
\usepackage{balance}
\usepackage{bm}
\usepackage{color}

\usepackage{times,mathptmx}
\usepackage{sectsty}
\usepackage{graphicx} 
\usepackage{lastpage}
\usepackage[format=plain,justification=justified,singlelinecheck=false,font={stretch=1.125,small,sf},labelfont=bf,labelsep=space]{caption}
\usepackage{float}
\usepackage{fancyhdr}
\usepackage{fnpos}
\usepackage[english]{babel}
\addto{\captionsenglish}{%
  
}
\usepackage{array}
\usepackage{droidsans}
\usepackage{charter}
\usepackage[T1]{fontenc}
\usepackage[usenames,dvipsnames]{xcolor}
\usepackage{setspace}
\usepackage[compact]{titlesec}
\usepackage{hyperref}

\usepackage{epstopdf}

\definecolor{cream}{RGB}{222,217,201}

\begin{document}

\pagestyle{fancy}
\thispagestyle{plain}
\fancypagestyle{plain}{

\fancyhead[C]{\includegraphics[width=18.5cm]{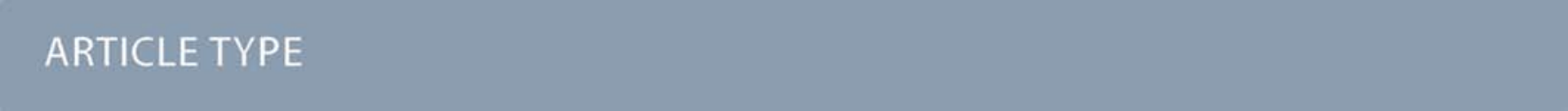}}
\fancyhead[L]{\hspace{0cm}\vspace{1.5cm}\includegraphics[height=30pt]{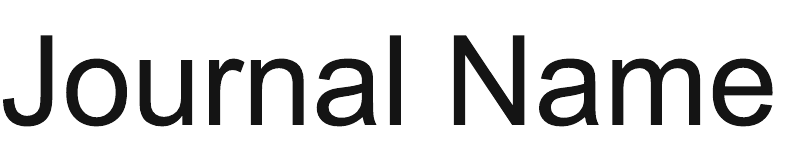}}
\fancyhead[R]{\hspace{0cm}\vspace{1.7cm}\includegraphics[height=55pt]{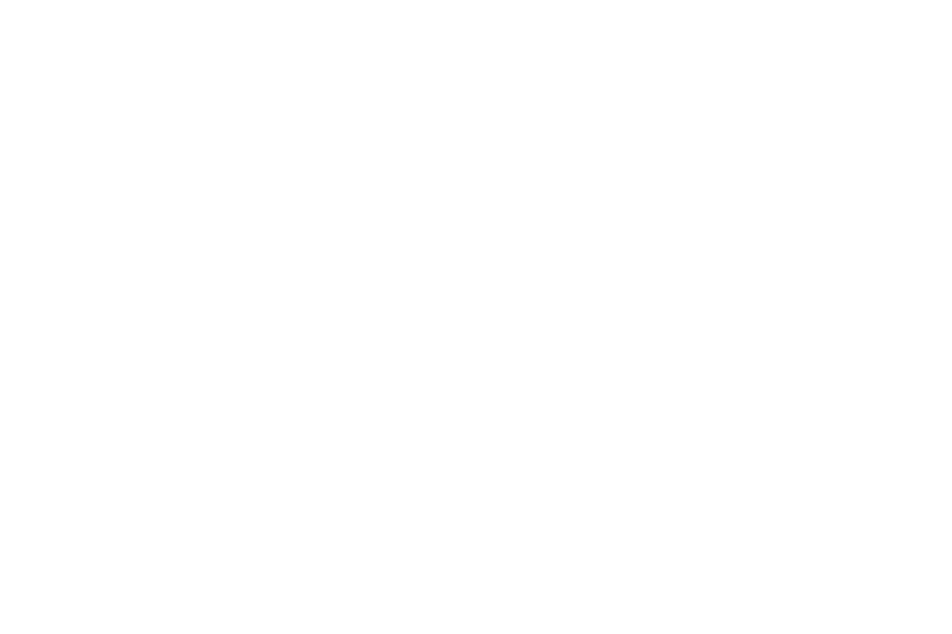}}
\renewcommand{\headrulewidth}{0pt}
}

\makeFNbottom
\makeatletter
\renewcommand\LARGE{\@setfontsize\LARGE{15pt}{17}}
\renewcommand\Large{\@setfontsize\Large{12pt}{14}}
\renewcommand\large{\@setfontsize\large{10pt}{12}}
\renewcommand\footnotesize{\@setfontsize\footnotesize{7pt}{10}}
\makeatother

\renewcommand{\thefootnote}{\fnsymbol{footnote}}
\renewcommand\footnoterule{\vspace*{1pt}%
\color{cream}\hrule width 3.5in height 0.4pt \color{black}\vspace*{5pt}} 
\setcounter{secnumdepth}{5}

\makeatletter 
\renewcommand\@biblabel[1]{#1}            
\renewcommand\@makefntext[1]%
{\noindent\makebox[0pt][r]{\@thefnmark\,}#1}
\makeatother 
\renewcommand{\figurename}{\small{Fig.}~}
\sectionfont{\sffamily\Large}
\subsectionfont{\normalsize}
\subsubsectionfont{\bf}
\setstretch{1.125} 
\setlength{\skip\footins}{0.8cm}
\setlength{\footnotesep}{0.25cm}
\setlength{\jot}{10pt}
\titlespacing*{\section}{0pt}{4pt}{4pt}
\titlespacing*{\subsection}{0pt}{15pt}{1pt}

\fancyfoot{}
\fancyfoot[LO,RE]{\vspace{-7.1pt}\includegraphics[height=9pt]{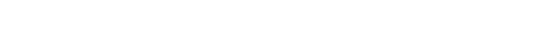}}
\fancyfoot[CO]{\vspace{-7.1pt}\hspace{13.2cm}\includegraphics{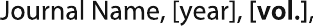}}
\fancyfoot[CE]{\vspace{-7.2pt}\hspace{-14.2cm}\includegraphics{RF}}
\fancyfoot[RO]{\footnotesize{\sffamily{1--\pageref{LastPage} ~\textbar  \hspace{2pt}\thepage}}}
\fancyfoot[LE]{\footnotesize{\sffamily{\thepage~\textbar\hspace{3.45cm} 1--\pageref{LastPage}}}}
\fancyhead{}
\renewcommand{\headrulewidth}{0pt} 
\renewcommand{\footrulewidth}{0pt}
\setlength{\arrayrulewidth}{1pt}
\setlength{\columnsep}{6.5mm}
\setlength\bibsep{1pt}

\makeatletter 
\newlength{\figrulesep} 
\setlength{\figrulesep}{0.5\textfloatsep} 

\newcommand{\topfigrule}{\vspace*{-1pt}%
\noindent{\color{cream}\rule[-\figrulesep]{\columnwidth}{1.5pt}} }

\newcommand{\botfigrule}{\vspace*{-2pt}%
\noindent{\color{cream}\rule[\figrulesep]{\columnwidth}{1.5pt}} }

\newcommand{\dblfigrule}{\vspace*{-1pt}%
\noindent{\color{cream}\rule[-\figrulesep]{\textwidth}{1.5pt}} }

\makeatother

\twocolumn[
  \begin{@twocolumnfalse}
\vspace{3cm}
\sffamily
\begin{tabular}{m{4.5cm} p{13.5cm} }

\includegraphics{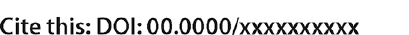} & \noindent\LARGE{\textbf{Propulsion and energetics of a minimal magnetic microswimmer$^\dag$}} \\
\vspace{0.3cm} & \vspace{0.3cm} \\

& \noindent\large{Carles Calero,\textit{$^{\ddag,a,b}$} Jos\'e Garc\'ia-Torres,\textit{$^{\ddag,a,b}$} Antonio Ortiz-Ambriz,\textit{$^{a,b}$} Francesc Sagu\'{e}s,\textit{$^{b,c}$} Ignacio Pagonabarraga,\textit{$^{a,d,e}$} and Pietro Tierno\textit{$^{\ast,a,b,e}$}} \\

\includegraphics{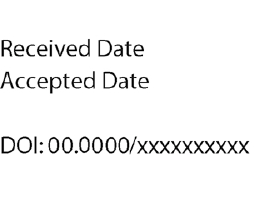} & \noindent\normalsize{In this manuscript we describe the realization of a minimal hybrid microswimmer, composed of
a ferromagnetic nanorod and a paramagnetic microsphere.
The unbounded pair is propelled in water upon application of 
a swinging magnetic field that induces 
a periodic relative movement of the two composing elements, 
where the nanorod rotates and slides on the surface of the paramagnetic sphere. When taken together, the processes of rotation and sliding describe a finite area in the 
parameter space which increases with the frequency of the applied field.  
We develop a theoretical approach and combine it with numerical simulations 
which allow us to understand the dynamics of the propeller and explain the experimental observations. Further, we 
demonstrate a reversal of the microswimmer velocity when varying the length of the nanorod, as predicted by the model. Finally we determine theoretically and in experiments the Lighthill's energetic efficiency of this minimal magnetic microswimmer.} \\

\end{tabular}

 \end{@twocolumnfalse} \vspace{0.6cm}

  ]

\renewcommand*\rmdefault{bch}\normalfont\upshape
\rmfamily
\section*{}
\vspace{-1cm}


\footnotetext{\textit{$^{a}$~Departament de F\'isica de la Mat\`eria Condensada, Universitat de Barcelona, Av. Diagonal 647, 08028, Barcelona, Spain. Tel: +34 9340 934034031; E-mail: ptierno@ub.edu}}
\footnotetext{\textit{$^{b}$~Institut de Nanoci\`{e}ncia i Nanotecnologia, Universitat de Barcelona, Barcelona, Spain. }}
\footnotetext{\textit{$^{c}$~Departament de Ci\`{e}ncia de Materials i Qu\'{i}mica F\'{i}sica, Universitat de Barcelona, Barcelona, Spain.}}
\footnotetext{\textit{$^{d}$~CECAM, Centre Europ\'een de Calcul Atomique et Mol\'eculaire, \'Ecole Polytechnique F\'ed\'erale de Lasuanne, Batochime, Avenue Forel 2, Lausanne, Switzerland}}
\footnotetext{\textit{$^{e}$~Universitat de Barcelona Institute of Complex Systems (UBICS), Universitat de Barcelona, Barcelona, Spain}}

\footnotetext{\dag~Electronic Supplementary Information (ESI) available: one file in .pdf format with more details on the theoretical model and two videoclips (.AVI format) as support of the experimental data.   See DOI: 00.0000/00000000.}


\section*{Introduction}
The ability to swim or propel at 
low Reynolds number ({\it Re}), 
when viscous effects dominate over inertial forces, 
characterizes many biological 
systems 
living in a fluid medium, such as cells or bacteria~\cite{Bra00,Ber04}. 
In this situation, the fluid flow can be considered as time-reversible,
and propulsion cannot be achieved 
with a simple reciprocal 
movement, namely 
a periodic sequence of back and forward displacements~\cite{Happ}.
Such necessary condition can be formulated in an equivalent way by stating that in a viscous,
Newtonian fluid propulsion can only be achieved 
when there are at least two independent
degrees of freedom able to describe a closed area in the parameter space~\cite{Pur97}.
The canonical example is the scallop, 
a macroscopic organism that is characterized by one 
degree of freedom, as it can only periodically open and close its shells~\cite{Lau11}.
Thus, the scallop provides a simple example of reciprocal motion,
and it would not swim at low {\it Re}. 
The same physical situation applies to 
artificial microswimmers, 
i. e. man-made micromachines 
that are designed to navigate in viscous fluids, and thus 
have to avoid reciprocal motion 
in order to translate in the medium.  

The current push to fabricate microscopic devices that can be propelled in fluid media is in part driven by  fundamental
motivations aimed at investigating the physics of microorganism motility and reproducing their essential features~\cite{Lau09}. 
But  there is also the technological promise to deliver a novel class of microscale devices capable of performing useful tasks in microfluidic chips~\cite{Tie082,San11} or vein networks~\cite{Pey13,Kim13,Bar16}. These motivations have lead to the realization of different prototypes based on the use of 
flexible or stiff parts that could be actuated by chemical reactions or external fields~\cite{Bec16}. 
In the latter case, some examples include the use of electric~\cite{FMa15,FMa152}, magnetic~\cite{Dre05,Zha09,Gho09,Sne09},
and optical fields~\cite{Vol11,Pal16} or ultrasound~\cite{Wan12,XuT14,Tai17}, to cite a few cases. 
In parallel, different prototypes have been 
realized via chemical reactions, such as Pt-Au nanorods  
that induce heterogeneous catalytic reactions
on their surface~\cite{Pax04}, Janus colloids~\cite{How07,Usp16}
or other reactive species~\cite{Gao2011,Palacci2013,Massana2018}.
In spite of all these achievements, experimental prototypes actuated by external magnetic fields demonstrated simplicity in design, as they require a finite number of well identified, independent, degrees of freedoms to propel in a fluid medium. Some examples include elementary stiff dumbbells~\cite{Tie08,Cheang2016}, triplets~\cite{Cheang2014}, chiral~\cite{Zha09,Gho09} or other planar achiral structures~\cite{Sachs2018,Tottori2018,Mirza2018,Cohen2019}.
Developing minimal microswimers is appealing for different reasons, even if their speed performance may be inferior to other prototypes which make use of flexible or helical parts. From a fundamental point of view, they allow to develop analytical models which can be used to understand the physical principle behind the interactions between few elements, to calculate analytically their efficiency or to clarify the role of hydrodynamics and the interaction with the dispersing medium. From the point of view of applications, minimal microswimmers can be easily optimized as their propulsion mechanism is based on the rotation/displacement of few elements and thus on a minimal number of degrees of freedom.

In our previous work~\cite{Calero2019}, we have realized experimentally a
hybrid magnetic microswimmer composed by 
a ferromagnetic nanorod and a paramagnetic microsphere. 
The couple assembled and propelled when it was actuated by a {\it swinging} magnetic field which performed periodic oscillations around a fixed axis and 
induced consecutive rotation and translation of the nanorod close 
to the surface of paramagnetic particle. These movements were uncoupled, and allow the displacement of the pair since they provided two
degrees of freedom. These are represented by two angles and   
describe a 
closed area in the parameter space.
In this article we generalize the previous study of 
the magnetic pair by formulating 
a theoretical approach which allows 
to characterize the displacement of the pair and 
the corresponding energetic efficiency. 
This model displays good agreement with 
numerical simulations which are performed in order to explore a 
range of parameters which are inaccessible to the experimental system. 
Further, we show experimentally that increasing the length of the 
nanorod
induces a reversion of the propulsion direction, another feature  captured with the theoretical model. 
We also demonstrate the scaling of the swimmer velocity with the area of the parameter space and the Lighthill's energetic efficiency of the microswimmer.

The article is organized as follow. First we
describe the experimental system and the technique used to measure the phase of the field in order to determine the energetic efficiency of the nanorod-particle pair. 
In the next section we explain the mechanism of motion of the pair, 
and then we introduce the theoretical model used to describe the experimental results. After that, we report the results 
of the theory and simulations 
where we identify and describe in detail the two independent degrees of freedom which enable the net translation of the propeller.
Then we describe how we measure the Lighthill's energetic efficiency of the microswimmer and discuss its dependencies on the parameters of the problem.  We conclude by stressing the
main results and provide a general outlook in view of this emergent, and rapidly developing research field.

\section*{Experimental part}
\subsection*{Materials and methods}
The micropropeller is composed of a magnetically assembled ferromagnetic nickel (Ni) nanorod and a paramagnetic microsphere. 
The Ni nanorods are synthesized by template-assisted
electrodeposition from a single electrolyte, $0.5 \rm{mol \, dm^{-3}}
NiCl_2$ solution (Sigma Aldrich), prepared with distilled
water treated with a Millipore (Milli Q system). The
electrosynthesis was conducted using a
microcomputer-controlled potentiostat/galvanostat Autolab
with PGSTAT30 equipment, GPES software and a three electrode-system. A polycarbonate (PC) membrane with pore
diameter $\sim 400 \rm{nm}$ (Merck-MilliPore) and sputter-coated
with a gold layer on one side to make it conductive is used as
the working electrode. The reference 
electrode is made of Ag/AgCl/KCl ($3\rm{mol \, dm^{-3}}$) while the counter electrode is a
platinum sheet. After synthesis, the Ni nanorods
are released from the membrane by first removing the gold
layer with a I$_2$/I$^-$ aqueous solution, and then by wet etching of the PC membrane in CHCl$_3$. Nanorods are then subsequently
washed with chloroform ($10$ times), chloroform-ethanol
mixtures ($3$ times), ethanol ($2$ times) and deionised water 
($5$ times). Finally, sodium dodecyl sulphate (Sigma Aldrich) is
added to disperse the nanorods. The typical length of the fabricated Ni
nanorods is $L =  3 \rm{\mu m}$. The
permanent moment of the nanorod is measured by following
its orientation under a static magnetic field, as described in
previous works~\cite{Mar16,Mart16}.
The value obtained for the magnetic moment of the ferromagnetic
rod is $m_n = 3.7\times10^{-11} \rm{Am^{-2}}$. 

\begin{figure}
\centering
\includegraphics[width=\columnwidth,keepaspectratio]{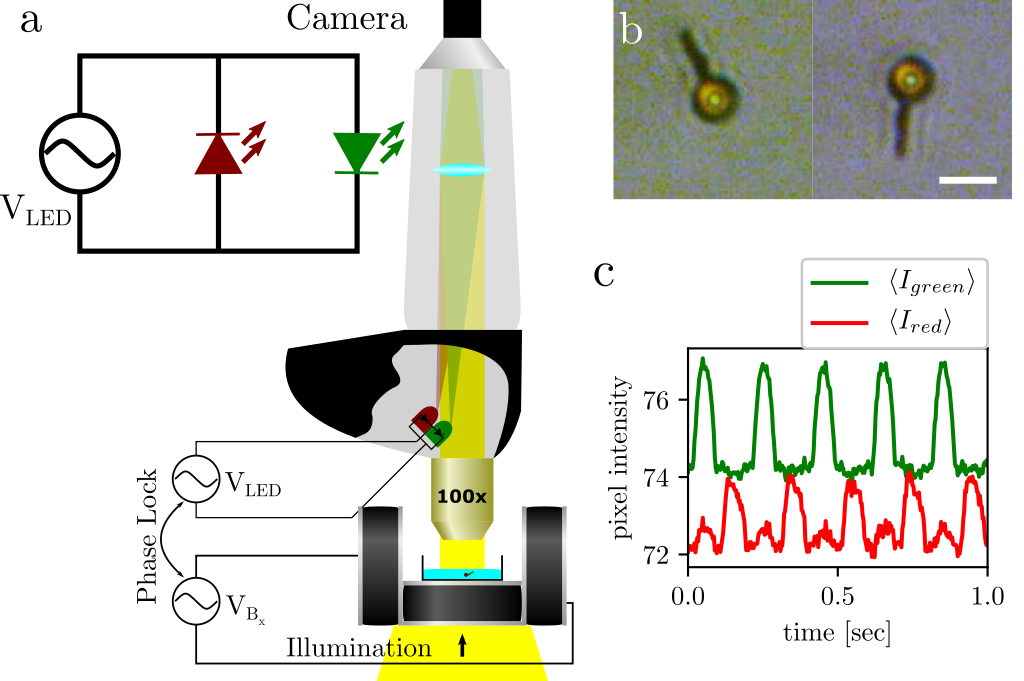}
\caption{(a) Diagram illustrating the experimental system
with the optical microscope and the set of coils arranged in the Helmholtz configuration and used to apply the swinging magnetic field. Inset shows the circuit used to connect the  two colored LEDs in an anti-parallel configuration to estimate the phase of the field at a given time (see main text). (b) The left panel shows an experimental image of the nanorod-particle propeller with the green LED on, and the right panel shows an image with the red LED on. The scale bar is $5\rm{\mu m}$. The difference in color is subtle, but can be appreciated more clearly in panel (c) where we plot the average value of the pixels of each channel versus time.}
\label{fig:ExperimentDiagram}
\end{figure}

The spherical colloids used are paramagnetic microspheres with radius $R = 1.5 \rm{\mu m}$, $\sim 15 \%$ iron oxide content and surface carboxylic groups (ProMag PMC3N, Bang Laboratories). 
The particles are characterized by a magnetic volume susceptibility equal to $\chi=0.21$, as measured in separate experiments~\cite{Jos18}. 
The particles and the nanorods are dispersed in highly deionized water (MilliQ, Millipore) and allowed to sediment above a glass substrate. 
While the fabrication yield of the Ni nanorods is high ($\sim 75 \%$), the final yield of the pair nanoro-microsphere is much lower. This is due to different reasons, as the fact that due to their permanent moment many nanorods tend to cluster and cannot be used to produce the microswimmer pair.

As shown in the schematic in Fig.1(a), the substrate is placed in the center of five orthogonal coils (for clarity only three coils are shown in the schematic), four of them arranged in the Helmholtz configuration on the stage of a light microscope (Eclipse Ni, Nikon). The latter is equipped with a Nikon $100\times$ objective with $1.3$NA. The coils are connected to a waveform generator (TGA1244, TTi) feeding a power amplifier (IMG AMP-1800). 
The particle dynamics are recorded with a CCD camera (scA640-74fc, Basler) working at $75$ frames per second (fps), with a CMOS camera (MQ003MG-CM, Ximea) working at $500$ fps, or in color at $325$ fps (acA640-750uc, Basler). 

\subsection*{Measurement of the phase of the field}
To measure the phase between the instantaneous value of the applied field and the orientation of the propeller, 
we modify the experimental set-up by introducing two light-emitting diode (LEDs) to the optical path, just above the observation objective, as shown in Fig.\ref{fig:ExperimentDiagram}(a). The two LEDs are connected in an anti-parallel configuration to an alternate current (AC) voltage source, which is produced by the same waveform generator that powers the magnetic coil system. We use a phase lock program to synchronize the oscillations coming from the two signals. In this configuration, the green LED emits light during the positive cycle
of the applied field, while the red LED emits on the negative one. The tube lens of the objective allows to distribute the colored light over the whole sample view. Even if the transmitted intensity appears as relatively small, it can be distinguished from the experimental image. From the color video in RGB format, we calculate the average value of all the pixels in the red and in the green channels as a function of time. An example of the time series for a $5\rm{Hz}$ video is shown in Fig. 
\ref{fig:ExperimentDiagram}(c). 
\begin{figure*}[t]
\begin{center}
\includegraphics*[width=\textwidth,keepaspectratio]{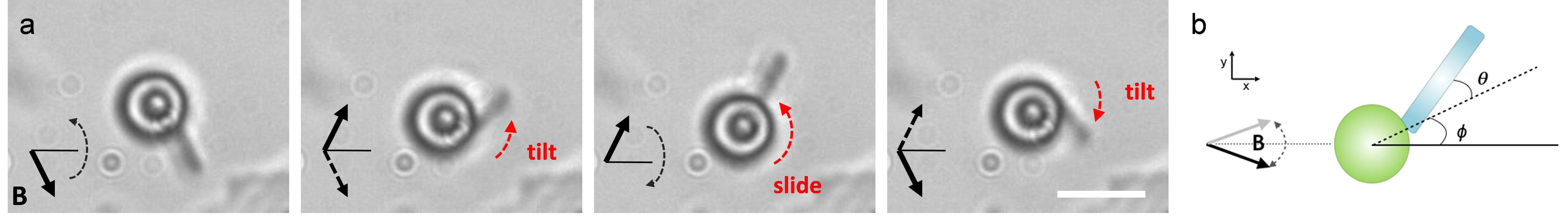}
\caption{(a) Sequence of experimental images showing 
one hybrid propeller composed by a paramagnetic colloid (radius $R=1.5 \rm{\mu m}$) and a Ni nanorod (length $L= 3 \rm{\mu m}$, diameter $D=400 \rm{nm}$). The pair is actuated by a swinging magnetic field with frequency $\nu = 20 \rm{Hz}$ and amplitudes $B_x= 2.15\rm{mT}$ and $B_y= 2.74\rm{mT}$.
In average the propeller moves towards left, with the particle in front of it. Scale bar is $5 \rm{\mu m}$, see VideoS1 in the Supporting Information. 
(b) Schematic picture of the system illustrating the definition of the two angles $\phi$ and $\theta$.}
\label{Fig2}
\end{center}
\end{figure*}
We then perform a least squares fit using the function 
$f(t) = A + \frac{B}{2} \left(\sin(2\pi ft + \phi)+
\left|\sin(2\pi ft + \phi)\right|\right)$ from which we extract the phase $\phi$, being $A$ and $B$ two amplitudes. The value of $\phi$ allows us to calculate an instantaneous value of the field for each frame. 
Further, we  track three points of the swimmer using the public program  ImageJ (National Institutes of Health). 
These points are the outermost tip of the nanorod, the point of contact between the nanorod and the colloidal particle, and the center of the colloidal particle. From these three points, we extract the relative angles, and using the instantaneous direction  of the applied field $\bf B$, we have all the information 
over the different degrees of freedom involved. 

\section*{Propulsion mechanism}
We start by describing the mechanism of motion of our assembled
microswimmer and how the nanorod-colloid pair reacts to the external field. 
Initially, the nanorod and the microsphere are separated, and can be approximated by a static magnetic field which induces attractive dipolar interactions. When the field is switched off, the pair can easily uncouple due to thermal fluctuations.
We induce propulsion by using a swinging 
magnetic field characterized by 
a static component 
of amplitude $B_x$ and an oscillating one perpendicular to it of
amplitude $B_y$ and frequency $\nu$:
\begin{equation}
\bm{B}\equiv (B_x,B_y \sin{(2 \pi \nu t)}) \, \, .
\label{appliedfield}
\end{equation}
Since the applied field is spatially homogeneous, i. e. it has no magnetic gradient, it produces no force on the particles. The field exerts only a torque to the nanorod, while it induces a dipole moment within the paramagnetic colloid.
The sequence of images in Fig.~\ref{Fig2}(a) results from fast video recordings of the hybrid propeller, which moves towards the left with the spherical particle in front of it. This sequence shows the orientation of the nanorod and its position relative to the spherical particle during one entire field cycle.
The sequence starts from a configuration where the permanent moment
of the nanorod ($\bm{m}_n$) and the induced moment of the particle ($\bm{m}_p$) are aligned 
along the field direction. Attractive dipolar interactions between the pair 
keep both elements together until the field changes direction. When the oscillating component of the field 
evolves and changes sign, $B_y \rightarrow -B_y$, we observe that the nanorod tilts following
the direction of the field by changing the orientation angle $\theta$,
defined in Fig.~\ref{Fig2}(b), while it keeps the contact point
with the particle surface approximately fixed. With a significant lag, the nanorod slides across the particle's surface varying the position of the contact point, which is characterized in terms 
of a second angle $\phi$ (see Fig.~\ref{Fig2}(b)). The translation of the 
nanorod on the spherical particle tends to align $\bm{m}_n$ with $\bm{m}_p$, minimizing 
the magnetic dipolar interaction energy of the couple. 
Another inversion of the oscillating component, $-B_y \rightarrow B_y$ repeats the cycle and the pair starts to advance 
with the paramagnetic particle in front of it. This sequence
of events showed in Fig.~\ref{Fig2}a is
also observed in the trajectories of our numerical simulations (see later). 

Since the the rotation and the sliding of the nanorod are mostly uncoupled, they provide the two independent degrees of freedom 
required to break the time-reversal symmetry of
the fluid flow, thus avoiding the Scallop's situation 
of reciprocal motion. 
Linking the nanorod to the particle, 
for example using $DNA$~\cite{Tie091}, would impede the 
relative sliding motion, thus reducing the degrees of freedom to one and making the motion reciprocal. 
Understanding the main ingredients of the propulsion 
mechanism is essential to justify the theoretical model that will
be detailed in the next section.   

Further we explore the average propulsion speed of the pair by fixing the static field to $B_x=2.15$mT and by varying both the amplitude of the oscillating field $B_y$ and the driving frequency $\nu$.
In particular, we vary $\nu \in [5,30]$Hz and $B_y\in [1.36,3.16]$mT and observe a maximum speed of $V=7\rm{\mu m/s}$ for the extreme values of both parameters~\cite{Calero2019}. 
Above $30$Hz, the nanorod becomes asynchronous with the swinging field and it cannot follow its fast oscillations, thus $30$Hz represent a critical (or step-out) frequency. As a consequence, the nanorod pins to the surface of the paramagnetic particles and there is no net motion of the pair.
The speed raises in a non linear manner with $\nu$ and $B_y$, 
but experimental limitations impeded us to explore the existence of a peak as a function of $\nu$,
due to the high range of frequency required. 
We note that in contrast with other prototypes as surface walkers~\cite{Mor08,Zha10,Sin10,Jos17} or rotors~\cite{Tie08,Bri13,Mar15,Del17}, 
that require the close proximity of a substrate to propel,
our case is completely independent from it. The substrate could 
have some effect on the microswimmer performance
but is not important for its propulsion mechanism.

\section*{Theoretical treatment}
We consider a pair composed of a spherical paramagnetic particle of radius $R$ and isotropic magnetic volume susceptibility $\chi$,
and a ferromagnetic rod of length $L$, diameter $D$, and a permanent moment a${\mathbf m}_n$. 
Both elements self-assemble due to their mutual magnetic dipolar interaction. The only external actuation is through the {\it swinging} magnetic field ${\bf B}$ given by Eq.~\ref{appliedfield}, which exerts a torque on the ferromagnetic rod. We assume that the magnetic field is sufficiently weak so that the paramagnetic particle  is in its linear response regime. Thus, the induced moment is given by ${\mathbf m}_p = \frac{4\pi}{3\mu_0}R^3 \chi {\mathbf B}$, being $\mu_0$ the permeability of vacuum. 
Due to the dimensions of the swimmer we assume that its dynamics are overdamped and governed by the hydrodynamic drag with the viscous fluid. 

\subsection*{Numerical simulations}
This theoretical picture of the microswimmer is implemented into a model whose dynamics under the actuating field are solved through numerical simulations. 

The dynamics of the swimmer are determined by magnetic, hydrodynamic, and steric interactions. To describe magnetic interactions we model the  paramagnetic colloid as a spherical particle with induced moment ${\mathbf m}_p$ located in its center. The shape and magnetization of the ferromagnetic rod are modeled using a group of $N$ equally spaced ferromagnetic beads of diameter $D$ and with permanent moment ${\bm m}_{n}/N$ directed along the axis of the rod. The length of the rod is kept fixed using a SHAKE algorithm. Under the actuating fields considered here we assume that the paramagnetic particle responds instantaneously, so the external field only exerts a torque on the ferromagnet, given by ${\bm \tau}_B =  {\bm m}_n \times {\bm B}$. This torque is implemented as an artificial force pair applied perpendicular to the axis of the rod. Additionally, the mutual magnetic interactions between the nanorod-sphere pair are modeled as a sum of dipolar interaction between the dipole of the paramagnetic sphere and the $N$ beads composing the ferromagnetic rod. 

The viscous fluid exerts a drag against the motion of the microswimmer. In addition, the independent motion of the two magnetic particles generates hydrodynamic flows which induce mutual hydrodynamic interactions.
In our description, the interaction of bead $i$ of the microswimmer with the viscous fluid is given by the hydrodynamic friction
force 
\begin{equation}
 {\bf F}_{H,i} = -\gamma_i({\bf v}_i - {\bf u}({\bf r}_i))\,,
\end{equation}
where $\gamma_i$ is the bead's friction coefficient, ${\bf v}_i$ its velocity, and ${\bf u}({\bf r}_i)$ 
is the induced fluid flow at the bead's position. 
We calculate the flow field ${\bf u}({\bf r})$ generated by the motion of the different components of the microswimmer in the far field regime.
Under this approximation we assimilate the hydrodynamic behavior of the beads composing the microswimmer
to that of point particles, which provides the flow field at point ${\bf r}_i$ 
\begin{equation}
 {\bf u}({\bf r}_i) = \frac{1}{8\pi \eta}\sum_{j} G({\bf r}_i;{\bf r}_j) \cdot {\bf F}_j\,.
\end{equation}
Here, ${\bf F}_j$ is the non-hydrodynamic force acting on particle $j$, $\eta$ is the viscosity of the fluid,  
and $G({\bf r}_i;{\bf r}_j)$ is the Oseen tensor.
To avoid overlaps of the particles composing the swimmer we also provide them with short ranged steric interactions through a Weeks-Chandler-Andersen potential \cite{WCA}. 

Under these interactions, the dynamics of the swimmer evolves following Newton's equations of motion. This is done by using a Verlet
algorithm adapted for cases with forces which depend on the velocity \cite{Bai09}.

\subsection*{Analytical perturbative solution}
We also solve analytically the dynamics of the self-assembled microswimmer under the action of a {\it swinging} field in the regime of small amplitude of the oscillating field, $\epsilon \equiv B_y/B_x \ll 1$. 
To accomplish that, we model the paramagnetic particle as a sphere with an induced magnetic moment ${\bm m}_{p}$ in its center, and the ferromagnet as a rod with a permanent moment ${\bm m}_{n}$ also located in its center. Dipolar attraction between the two magnetic particles keeps the tip of the rod on the surface of the paramagnetic sphere, a point which is identified as a virtual joint. The dynamical equations for the virtual joint in the overdamped regime are given by~\cite{Emi16}
\begin{eqnarray}
\label{Eq:dynamics}
 \mathbf{f}_r + \mathbf{f}_s &=& 0 \nonumber \\
 \mathbf{M}_r +  \mathbf{\tau}_B + \mathbf{r}_r \times \mathbf{f}_r +  \mathbf{r}_r \times \mathbf{f}_r & = & 0 \\
 \mathbf{r}_s \times \mathbf{f}_s - \mathbf{\tau}_{dip} &=& 0 \, \, \, \, , \nonumber 
\end{eqnarray}
where $\mathbf{f}_r$, $\mathbf{M}_r$ are the forces and torques exerted by the fluid on the rod, and $\mathbf{f}_s$ is the viscous drag force of the fluid on the spherical particle. $\mathbf{r}_r$ and $\mathbf{r}_s$ are the vectors from the joint to the center of the rod and sphere. Neglecting hydrodynamic interactions between the sphere and the rod, we approximate $\mathbf{f}_r$ and $\mathbf{M}_r$ with the help of slender body theory as
\begin{eqnarray}
 \mathbf{f}_r & = & -\frac{2\pi \eta L}{\gamma} (\mathbf{v}_r \cdot \hat{\mathbf{t}})\hat{\mathbf{t}} -  \frac{4\pi \eta L}{\gamma} (\mathbf{v}_r \cdot \hat{\mathbf{n}})\hat{\mathbf{n}} \label{Eq:slender_body}\\ 
 \mathbf{M}_r & = & -\frac{\pi \eta L^3}{3\gamma}\mathbf{\omega}_r \, \, \, . 
\end{eqnarray}
Here, $\eta$ is the viscosity of the fluid, $\gamma = \log(L/D)$, $\mathbf{v}_r$ and $\omega_r$ are the linear and angular velocities of the rod, respectively.  In Eq.~\ref{Eq:slender_body}, $\hat{\mathbf{t}}$ ($\hat{\mathbf{n}}$) denotes the direction parallel (perpendicular) to the axis of the rod. In eqs.~\ref{Eq:dynamics},
$\mathbf{f}_s$ is given by Stokes law $\mathbf{f}_s = -6\pi \eta R \mathbf{v}_s$, where $\mathbf{v}_s$ is the linear velocity of the spherical particle. In this model we assume that the spherical particle is a perfect paramagnet, so there is no external torque exerted on it and as a result it does not rotate. 

In Eq.~\ref{Eq:dynamics}, $\mathbf{\tau}_{dip}$ is the internal torque exerted by the dipolar forces about the joint, which defines the shape of the swimmer and can be written as
\begin{equation}
 \tau_{dip} = \frac{3\mu_0}{4\pi r^5} \mathbf{r}\times [ \mathbf{m}_n(\mathbf{r} \cdot \mathbf{m}_p) + \mathbf{m}_p(\mathbf{r}\cdot \mathbf{m}_n) ] \, \, \, ,
\end{equation}
where $\mathbf{r}$ is the vector joining the centers of the paramagnetic and ferromagnetic particles. 

\section*{Results}

\subsection*{Velocity of the swimmer}
The relevant coordinates of the problem are the coordinates of the joint on the plane of motion $(x, y)$ and the angles $\theta$ and $\phi$ defined by the position of the joint on the surface of the spherical particle and the direction of the rod, see Fig.~\ref{Fig2}b. The problem is controlled by three dimensionless parameters, namely the ratio between the sizes of the particles, $\delta = {L}/{R}$, the volume paramagnetic susceptibility $\chi$ and $\alpha_{ind} = {\mu_0 m_n}/({R^3 B_x})$, which is a measure of the importance of the magnetic field induced by the ferromagnet on the paramagnet with respect to the external field.  

Following the work by Gutman and Or~\cite{Emi16}, we solve the steady state dynamics of the propeller with the transfer function formalism using a perturbative approach under the condition of small oscillating transverse field, $\epsilon \ll 1$. The leading order expression for the velocity of the propeller $V$ is given by
\begin{equation}\label{Eq:velocity}
 V/V_0 = \epsilon^2 \frac{ b_1 (\nu/\nu_0)^2}{a_4(\nu/\nu_0)^4 + a_2(\nu/\nu_0)^2 + a_0} + \mathcal{O}(\epsilon^4)  \,,
\end{equation}
where $V_0 = R \nu_0$, and $\nu_0 = \frac{9 B_x m_n}{\pi \eta L^3}$ is a characteristic frequency of the problem.
$b_1, a_0, a_2, a_4$ are coefficients which depend on the parameters $\delta, \chi, \alpha_{ind}$ and are provided in a script in the Supporting Information. In our approach, collecting analytical expressions for higher order terms in the relative amplitude $\epsilon$ becomes unmanageable.

\begin{figure}[t]
\begin{center}
\includegraphics*[width=\columnwidth]{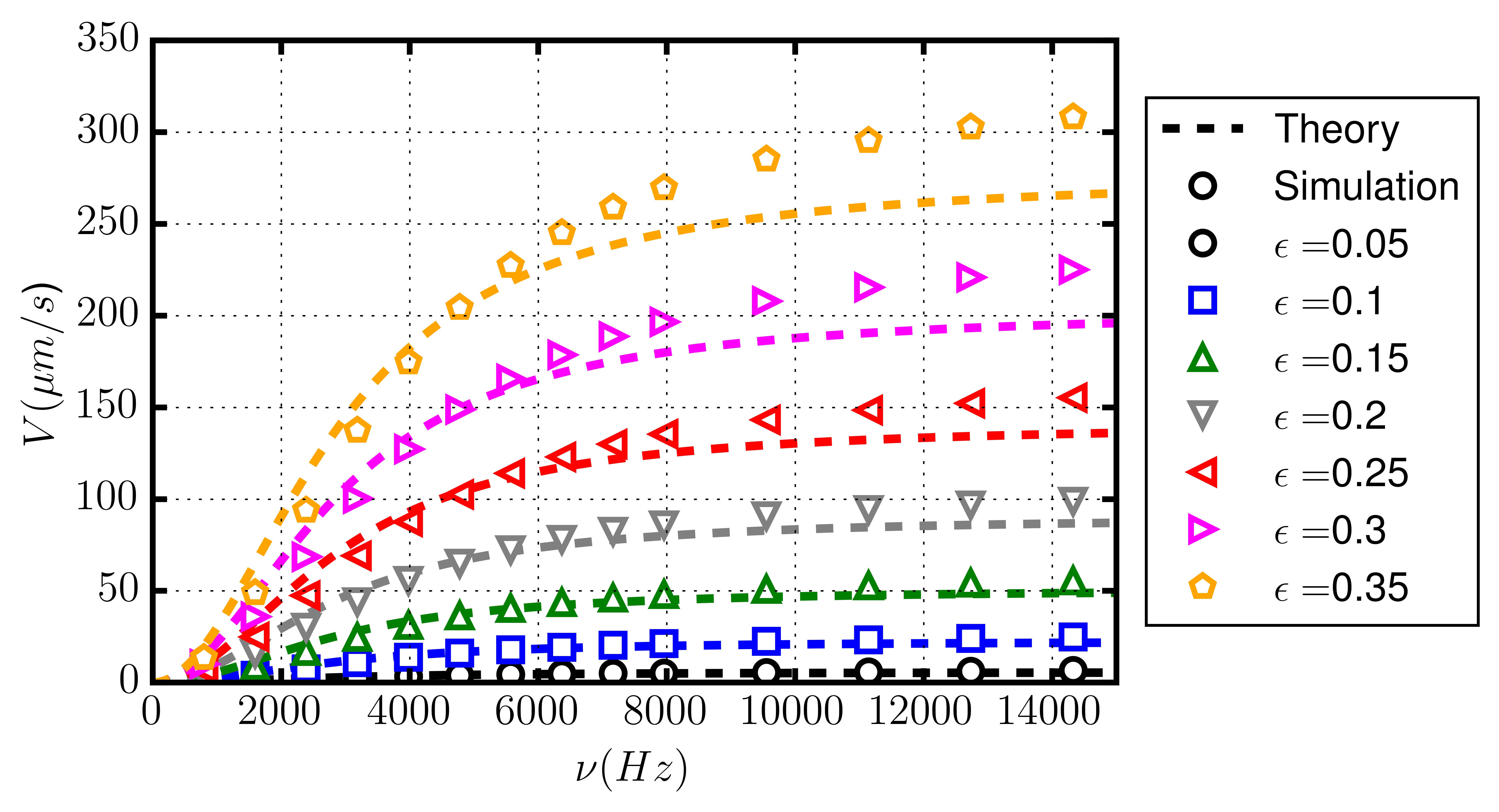}
\caption{Velocity $V$ of the microswimmer 
versus driving frequency $\nu$.  Symbols correspond to the results obtained from 
numerical simulations. Dashed lines represent the results from the analytical model.}
\label{Fig:velocity}
\end{center}
\end{figure}

The dependence of $V$ on the frequency for typical values of the field ratio $\epsilon$ is shown in Fig.~\ref{Fig:velocity}. Along with the theoretical curves we plot also results from the numerical simulations (dashed points) over a wide range of driving frequencies, in order to display the complete functional form. As shown
previously~\cite{Calero2019}, within the range of experimentally accessible frequencies $\nu \leq 30$Hz the velocity exhibits a good agreement with the experimental data.  
Further we note that Eq.~\ref{Eq:velocity} exhibits a maximum at frequency
\begin{equation}
 \nu_{V}^{max}/\nu_0 = \left(\frac{a_0}{a_4} \right)^{1/4} \, .
\end{equation}

The model is based on the assumption of the existence of a contact point between the ferromagnetic rod and the paramagnetic sphere, which restricts the range of parameters to cases where the dipolar attraction between the two magnetic particles is large enough. In addition, the expressions are derived under the condition of small oscillating transverse fields, $ \epsilon \ll 1$. Within this regime, the analytical model provides a good quantitative account of the velocity of the microswimmer obtained with numerical simulations in the whole range of frequencies, as shown in Fig.~\ref{Fig:velocity}. Note that even for values of $\epsilon$ of the order of $0.35$, we obtain a good agreement between the analytical formula Eq.~\ref{Eq:velocity} and the results from numerical simulations.

\begin{figure}[t]
\begin{center}
\includegraphics[width=\columnwidth,keepaspectratio]{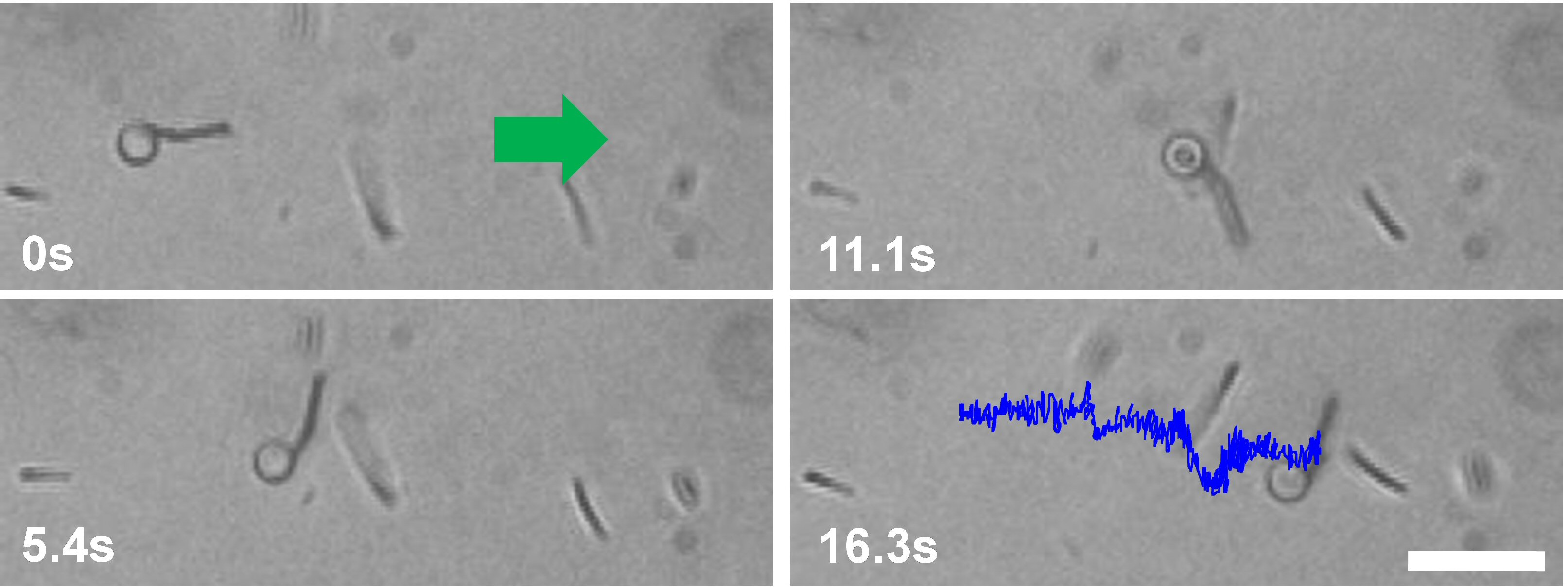}
\caption{Sequence of microscope images showing the propulsion 
of the hybrid microswimmer 
with a long nanorod in front of it, length $L \sim 6 \rm{\mu m}$.
The green arrow in the first image denotes the 
direction of motion. 
The external field parameters are
$B_x= 2.15 \rm{mT}$, $B_y= 1.3 \rm{mT}$
and frequency $\nu = 10 \rm{Hz}$.
The tracking of the center of velocity of the pair is superimposed to the bottom image.
The  scale bar is $10\rm{\mu m}$, see VideoS2 in the Supporting Information.}
\label{Fig:reversal}
\end{center}
\end{figure}

\subsubsection*{Inversion of direction of motion.}
The coefficient $b_1$ in Eqs.~(\ref{Eq:velocity}) can be written as 

\begin{eqnarray}
 b_1 &=& A(\delta, \alpha_{ind})\big\{3\pi\gamma\left[\delta^4 + 8\delta^3 + 24\delta^2 + 32\delta + 16 \right] \nonumber \\&-& {8\pi\chi}\gamma \left[\delta^3/\gamma + 6\delta^2 + 18\delta + 18 \right]\big\} \, \, \, .
\end{eqnarray}
This indicates the existence of a threshold $\chi*$ which determines the direction of motion of the microswimmer pair
\begin{equation}
 \chi* = \frac{3\left(\delta^4 + 8\delta^3 + 24\delta^2 + 32\delta + 16 \right)}{8\left(\delta^3/\gamma + 6\delta^2 + 18\delta + 18 \right)} \, \, \, .
\end{equation}
For $\chi < \chi*$ the velocity of the propeller is positive, with the paramagnetic particle in front of the pair and moving along the x axis, whereas for $\chi > \chi*$ it becomes negative, with the pair moving along the $x<0$ direction with the nanorod in front of it. Thus, for a fixed diameter of the magnetic particle, the direction of motion can be reversed by considering paramagnetic particles with different magnetic susceptibility $\chi$. Alternatively, for a fixed $\chi$ the ratio $\delta = L/R$ controls the value of $\chi*$ and thus can also determine the sign of the propulsion velocity. 

We demonstrate experimentally the possibility of inverting the direction of motion by using longer nanorod, as shown in Fig.4. 
Indeed, while the nanorod used previously, with $L\sim 3 \rm{\mu m}$,  moves with the spherical 
paramagnetic particle at the front, we observe that the direction of propulsion inverts when using a longer nanorod with $L\sim 6 \rm{\mu m}$ 
and the pair moves with the nanorod at the front of it.
This behavior provides a further functionality to our microswimmer, 
namely the possibility to control the direction of motion also by tuning 
the geometrical parameters, and not only the chirality of the applied field. This effect could be eventually used in a suspension of several hybrid microswimmer, to realize a selective sorting by inducing an opposite motion. The sorting effect could be used to select a population of microswimmers characterized by nanorods with similar lengths, which then could be separated from the particles by simply switching off the applied field.

\subsection*{Relation between velocity of microswimmer and the cyclical deformations}

\begin{figure}[ht!]
\begin{center}
\includegraphics[width=0.95\columnwidth]{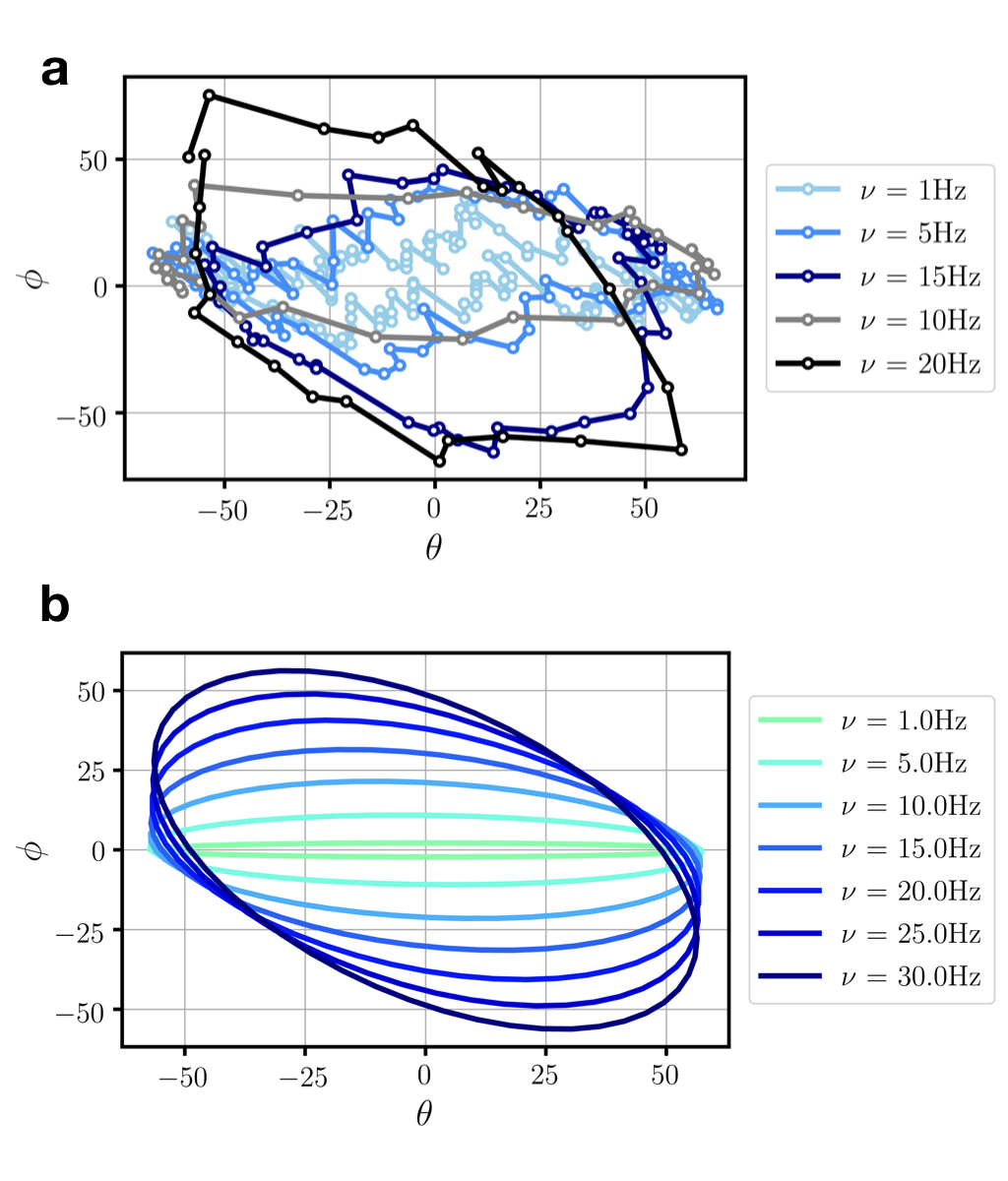}
\caption{(a,b) Cycles in the [$\theta, \phi$] plane of the sphere-rod swimmer at different frequencies from experiments taken from Ref.~\cite{Calero2019} (a) and from the theoretical model (b).}
\label{Fig:area}
\end{center}
\end{figure}

The propulsion of the microswimmer is allowed because it undergoes periodic non-reciprocal deformations, defined by the angular coordinates $\phi, \theta$. 
The perturbative solution to the theoretical model provides expressions for the steady-state evolution of the coordinates $(\phi,\theta)$ which define the deformation of the microswimmer. To first
 order in $\epsilon$, they are given by
 \begin{eqnarray}\label{Eq:angles}
  \phi(t)/\epsilon &=& A_{\phi}\sin(2\pi \nu t) + B_{\phi}\cos(2\pi \nu t) + \mathcal{O}(\epsilon) \nonumber \\
  \theta(t)/\epsilon &=& A_{\theta}\sin(2\pi \nu t) + B_{\theta}\cos(2\pi \nu t) + \mathcal{O}(\epsilon)\,.
 \end{eqnarray}
The coefficients $A_{\phi}, B_{\phi}, A_{\theta}, B_{\theta}$ (given in a script in the Supporting Information) are functions of $\nu$ and of the dimensionless parameters which define the problem $\delta, \alpha_{ind}, \chi$. 
Due to the simplicity of our magnetic swimmer, we can track in experiment the dynamics of such deformations. In Fig.~\ref{Fig:area} we show the trajectories in the parameter space $(\phi(t),\theta(t))$ obtained in experiments and from the analytical model for different frequencies. The cyclical trajectories enclose a finite area which evidences the non-reciprocity of the motion.

In the following we demonstrate an approximate relation between the average speed of our microswimmer and the trajectory of its cyclical deformations. 
The dynamics within a cycle of the magnetic field is governed by the rotation of the ferromagnetic nanorod following 
the direction of the external magnetic field. In addition, the ferromagnetic rod moves on the surface of the paramagnetic 
spherical particle to minimize the dipolar interaction. At low {\it Re} the velocity of the swimmer is proportional to the forces and torques externally applied to the object through the mobility matrix, ${\bf V} = {\bf M} {\bf F}$, where ${\bf F}$ contains the external forces and torques applied. In our case, there is only an external magnetic torque applied on the ferromagnetic particle along the z-direction perpendicular to the confining plane, $\tau_B = m B \sin(\alpha - \theta_B)$. The average velocity of the swimmer along the x-direction over a cycle of the external field can be expressed as
\begin{equation}\label{Eq:Vx}
 V^x = \nu \int_0^{1/\nu}dt M_{TR}^{xz}(t)\tau_B^z(t) \, \, \, ,
\end{equation}
where $M_{TR}$ is the part of the mobility matrix that couples the application of a torque with the translational motion. The external magnetic field induces a rotation of the ferromagnetic nanorod, whose angular velocity can be approximated to $\dot{\alpha} \approx -\tau_B^z/\xi_R$ under the assumption that dipolar induction only determines the position on the surface of the spherical particle. Here, $\alpha = \phi + \theta$ is the angle between the magnetic moment of the ferromagnet nanorod and the x-axis, $\xi_R$ is the rotational drag coefficient, which is
$\xi_R = \pi \eta L^3/(4 \log(L/a))$ for a rod of length $L$ and width $a$. As a result of the rotation of the nanorod, a flow is generated which drags the paramagnetic particle. This flow can be approximated by the flow induced by a point rotlet located at the center of the ferromagnetic rod. Under this approximation, and considering that the drag of the swimmer is mainly determined by the spherical particle, the mobility matrix becomes
\begin{equation}
M_{TR}^{xz}(t) \approx -\frac{\sin(\theta)}{2\pi\eta (D+L)^2}\,.
\end{equation}
Consequently, Eq.~\ref{Eq:Vx} becomes
\begin{equation}\label{Eq:Vx2}
V^x \propto \nu \left\{\int_{-\alpha_M}^{\alpha_M}d\alpha  \sin\theta  + \int_{\alpha_M}^{-\alpha_M}d\alpha  \sin\theta \right\} \equiv \nu A_X\,,
\end{equation}
where $\alpha_M = \arctan(B_{y}/B_{x})$ is the maximum angle between the nanorod and the horizontal, x-axis. Eq.~\ref{Eq:Vx2} tells us that the averaged velocity in a cycle is given by the enclosed area that is swept in a full cycle in the [$\sin{\theta}, \alpha$] space. A similar expression is found for the y-component of the average velocity, $V^y \propto \nu \left\{\int_{-\alpha_M}^{\alpha_M}d\alpha  \cos\theta  + \int_{\alpha_M}^{-\alpha_M}d\alpha  \cos\theta \right\} \equiv \nu A_Y$.

\begin{figure}[ht!]
\begin{center}
\includegraphics[width=0.9\columnwidth]{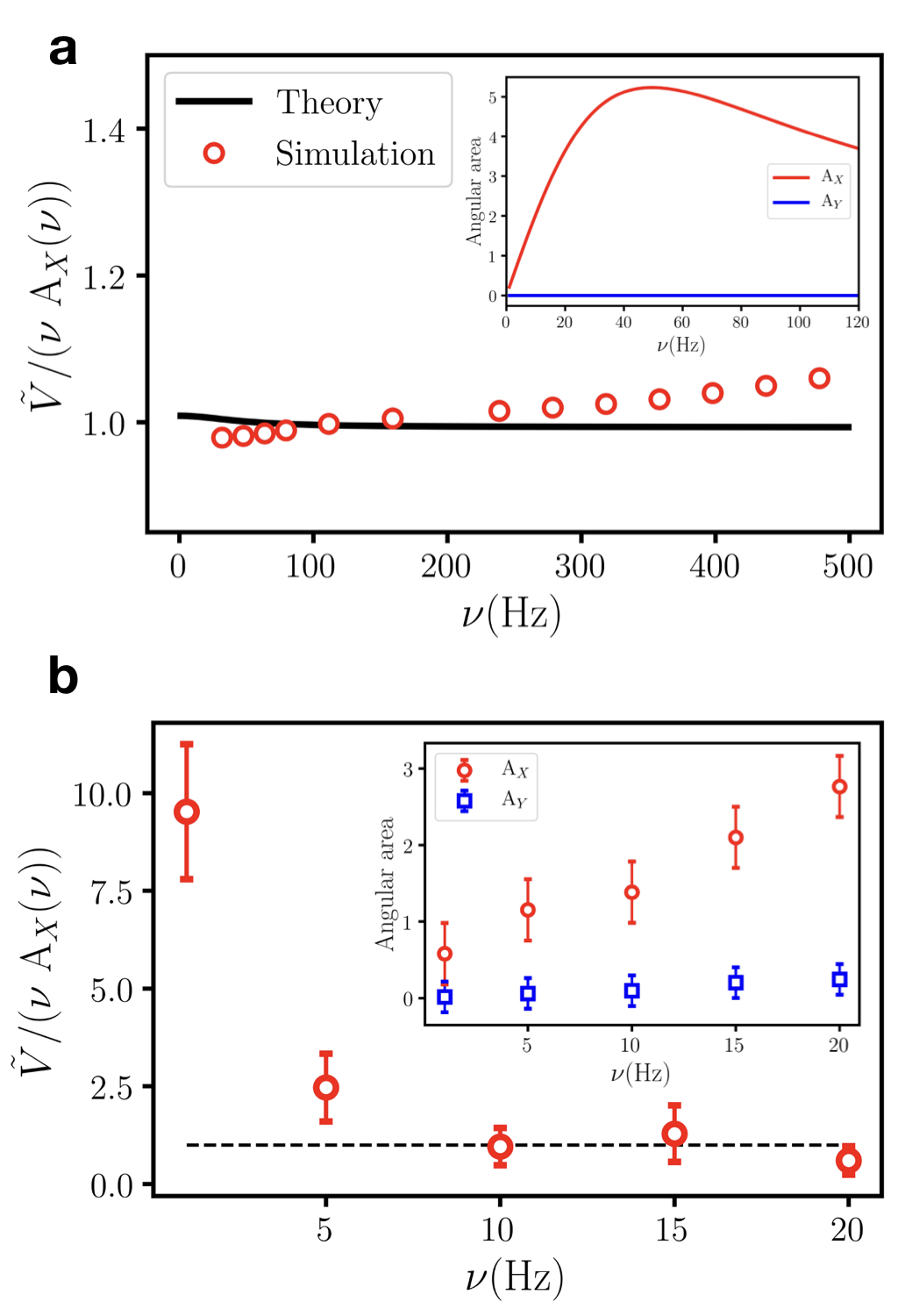}
\caption{(a,b) Scaling of the swimmer velocity with $\nu A_X$ ($A_X$ is the area enclosed in [$\sin{\theta}, \alpha$] space in a full cycle) for (a) numerical simulations 
(dots) and the perturbative solution of the theoretical
model (lines), and (b) experiments. The insets show the dependence
of $A_X$ and $A_Y$ on the driving frequency for (a) the
theoretical model, (b) experiments. 
The error bars in the experimental data are obtained from the statistical analysis of different experiments.}
      \label{Fig:velarea}
   \end{center}
 \end{figure}

In Fig.~\ref{Fig:velarea}a we show the normalized speed of the swimmer scaled by the driving frequency and the area enclosed by the [$\sin{\theta}, \alpha$] trajectory during a whole cycle, $\nu A_X$, as obtained in numerical simulations and with the perturbative solution of the theoretical model. In both approaches, $V/(\nu A_X)$ is almost constant regardless of the frequency of actuation. In the inset of Fig.~\ref{Fig:velarea}(a) we show the dependence of $A_X$ and $A_Y$ on the frequency. Note that due to symmetry, the integration through a whole cycle yields $A_Y \approx 0$, consistent with the zero velocity measured along the y-direction.
A similar analysis has been done for the experimental results, shown in Fig.~\ref{Fig:velarea}(b). 
The dependence of the angular areas $A_X$ and $A_Y$ on the frequency of actuation agrees well with the values obtained from the theoretical model (see inset in Fig.~\ref{Fig:velarea}b). The scaled speed, however, oscillates around a constant value only for the highest frequencies, with large deviations for $\nu = 1$Hz and $\nu = 5$Hz. At such low frequencies both the velocity and the area enclosed in the [$\sin{\theta}, \alpha$] space are very small numbers, and their exact determination becomes more difficult due to the presence of different sources of error. For example, at low speed thermal fluctuations may influence the measurement of the speed of the pair, or the close proximity of the substrate may introduce a further drag to the translational movement. When the area is small, the sliding motion of the nanorod tip may be affected by the roughness of the particle surface or the presence there of polymeric groups which may introduce additional steric interactions between the two elements. 

\subsection*{Determination of the Lighthill's energetic efficiency}
The Lighthill's efficiency $e$ 
compares the external power $\langle \Phi \rangle$ which is needed to induce a mean velocity ${V}$ in a medium of viscosity $\eta$, to the 
power required to rigidly drag the swimmer at the same speed with an external force $F_{drag}$~\cite{Lig75,Pu97},
\begin{equation}
e = \frac{F_{drag}{V}}{\langle \Phi \rangle}
\end{equation}
This parameter is a standard measure for the efficiency of many microswimmers, and has been employed in the past in different
theoretical works~\cite{Sha89,Lau09,Ost11} to analyze
the performance of simple
artificial designs like three-link flagella~\cite{Bec03,Pas12,Wie16},
three-sphere swimmers~\cite{Ram08},
squirmers~\cite{Ishi14}, neck-like propellers~\cite{Raz08} and
undulating magnetic systems~\cite{Emi16}.


\subsubsection*{Theoretical expression for the efficiency}
In our case, the Lighthill efficiency of the microswimmer for a given actuation is given by $e = \xi V^2/\bar{P}_B$, where $\xi$ is the drag coefficient of the propeller moving as a rigid body, $V$ its steady state average velocity, and $\bar{P}_B$ is the average power supplied by the external magnetic field. The power exerted on the swimmer is due to the action of the external field on the ferromagnetic nanorod
\begin{equation}\label{Eq:Power}
 P_B(t) =  \tau_B(t)\, \dot{\alpha} \, \, \, .
\end{equation}
Here, $\tau_B$ is the instantaneous torque exerted by the field on the rod, $\tau_B(t) = {\bf m}_n(t)\times{\bf B} (t)$, and $\dot{\alpha}$ is the angular velocity of the rod. The average power in one period of the magnetic field is givenby:
\begin{equation}\label{Eq:PB}
 \bar{P}_B = \nu \int_0^{1/\nu} P_B(t) dt \, \, \,.
\end{equation}
The instantaneous power $P_B(t)$ can be calculated using the perturbative solutions for the  evolution of the angles of deformation $(\phi,\theta)$, see Eqs.~\ref{Eq:angles}. Using Eq.~\ref{Eq:PB} and the analytic expression for the steady state average velocity of the microswimmer (Eq.~\ref{Eq:velocity}) we find that, to leading order in $\epsilon$ 
\begin{equation}\label{Eq:efftheor}
e = \epsilon^2 \frac{ n_2 (\nu/\nu_0)^2}{d_6(\nu/\nu_0)^6 + d_4(\nu/\nu_0)^4 + d_2(\nu/\nu_0)^2 + d_0} + \mathcal{O}(\epsilon^4)\,.
\end{equation}
Here, $d_0, d_2, d_4, d_6$ are coefficients which depend on the parameters of the problem $\delta, \chi, \alpha_{ind}$ and are provided in a script in the Supporting Information. 

\begin{figure}[t]
\begin{center}
\includegraphics*[width=\columnwidth]{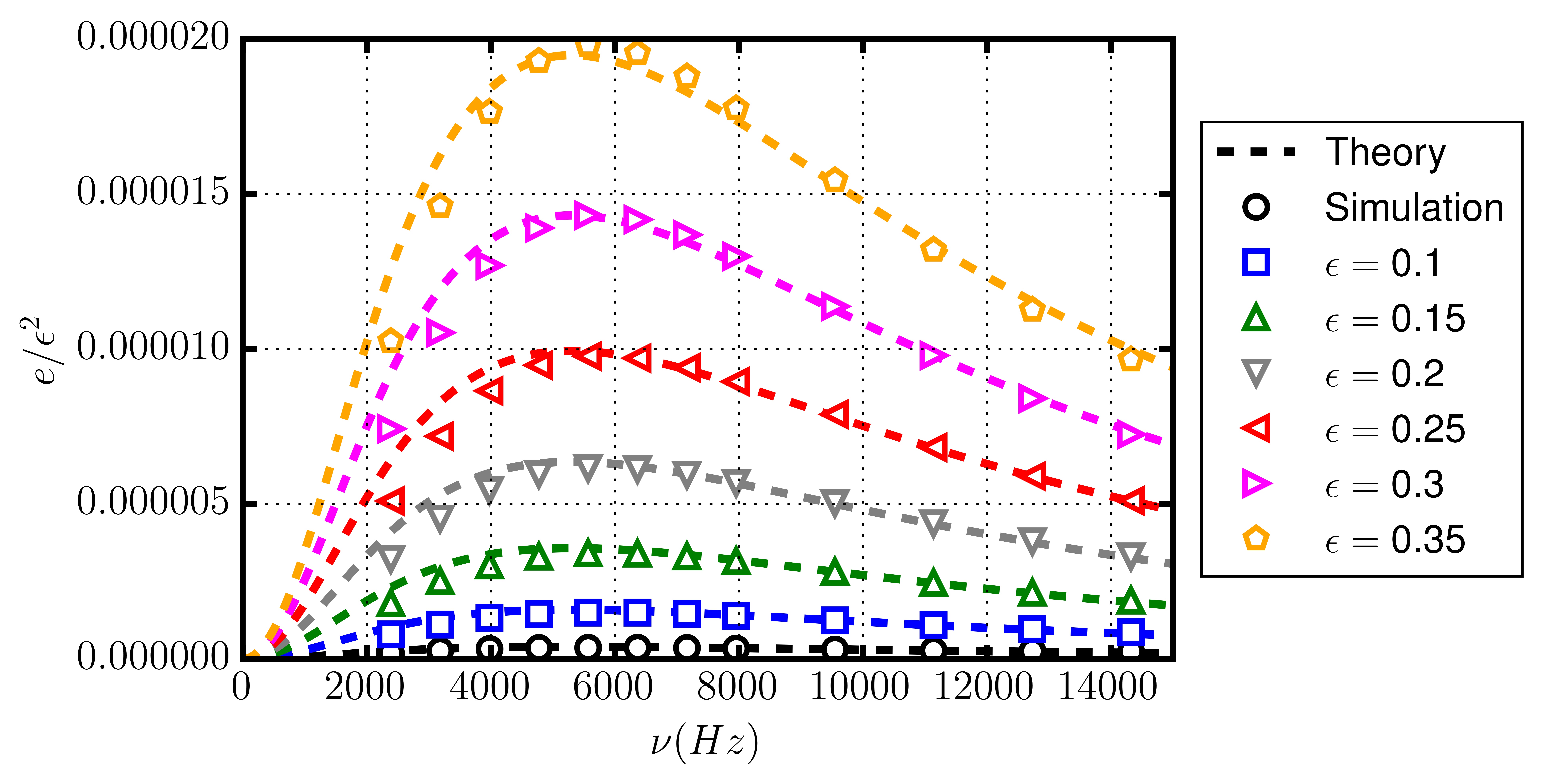}
\caption{Efficiency of the microswimmer 
versus driving frequency $\nu$. Symbols correspond to the results obtained from 
numerical simulations. Dashed lines represent the results from the analytical model.}
\label{Fig:efficiency}
\end{center}
\end{figure}

In Fig.~\ref{Fig:efficiency} we show the dependence of the efficiency on the frequency of the actuating field for typical parameters of the microswimmer. In addition, in Fig.~\ref{Fig:efficiency} we compare the results from Eq.~\ref{Eq:efftheor} and the efficiency of the microswimmer obtained from our numerical simulations, with excellent agreement in the whole range of frequencies. 


\subsubsection*{Experimental measurement}
\begin{figure}[t]
\begin{center}
\includegraphics*[width=\columnwidth]{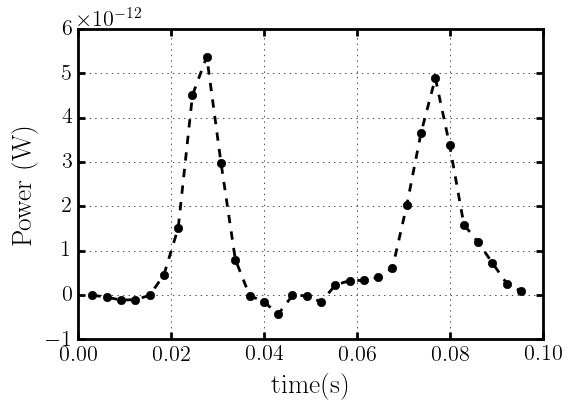}
\caption{Power at each instant of time for a cycle of the magnetic field with frequency $\nu = 10$Hz, $B_x = 2.15$mT and $B_y = 1.37$mT. }
\label{Fig:Power}
\end{center}
\end{figure}

In experiment we can directly measure the efficiency of the swimmer thanks to the precise tracking of the position of the microswimmer elements over time. To obtain $\bar{P}_B$ we record the dynamics of the swimmer with the help of a high-speed camera, taking images at 380 fps. With those images we track the position and orientation of the rod, which allows us to determine its angular velocity at each instant of time. To know the phase of the transverse component of the field at a given time from the analysis of the images we couple the signal of the alternating current current flowing through the Helmholtz coils to a pair of light emitting diodes.  
The instantaneous power exerted by the external field is obtained by using Eq.~\ref{Eq:Power} and the average power is computed by numerically integrating Eq.~\ref{Eq:PB}, see Fig.~\ref{Fig:Power}. The final average value for $\bar{P}_B$ is obtained by taking the average over 5 to 8 field cycles. 

The power needed to rigidly drag the swimmer is also obtained completely from parallel experiments. We use a single coil to generate a magnetic field, whose gradient decays with distance to the coil over a few centimeters. By measuring the velocity induced by the gradient at different distances we determine the external force, and thus the power, needed to drag the swimmer at a given velocity $V$ by interpolation. Independently, we can estimate the drag coefficient of the swimmer $\xi$ from its dimensions, and calculate the power needed to move it at a velocity $V$, $P = \xi V^2$. Both methods provide similar results, consistent within the uncertainty of the quantities. 

With this procedure we have directly measured the efficiency of our microswimmer for a frequency of $\nu = 10$Hz, a longitudinal magnetic field $B_x = 2.15$mT and a transverse amplitude of the magnetic field $B_y = 1.37$mT. Under these conditions we obtain an experimental value for the efficiency of $e^{exp} = (2.1 \pm 0.5)\times10^{-8}$. This value is significantly smaller than the efficiency obtained from theory and numerical simulations, which results in $e^{sim} = 2.9\times10^{-7}$. Such discrepancy can be attributed to a number of contributions which our theoretical model does not capture. Although in our case the propulsion mechanism is not affected by the presence of a bounding wall, the direct interaction with the wall due to gravity and its effects on the rheology of the fluid can significantly affect the magnitude of propulsion and hence the value of the Lighthill efficiency. In addition, our assumption of perfect paramagnet is a valid approximation to capture the essence of the propulsion mechanism but might induce significant quantitative discrepancies in the calculation of the input power. Indeed, the paramagnetic particle is composed of magnetic domains in a non magnetic matrix and can present residual magnetic anisotropy. As a consequence, the field can induce the rotation of the spherical particle and thus the dissipation of energy without contributing to the steady state velocity of the swimmer. Finally, the far field approximation assumed in our theoretical treatment of hydrodynamic interactions may also contribute to the theoretical overestimation of the microswimmer's efficiency. 

\section*{Conclusions}
In this article we have realized 
and fully characterized a minimal  
self-assembled microswimmer 
composed by a paramagnetic microsphere and 
a ferromagnetic nanorod. 
We have developed a theoretical model combined with numerical simulation to understand the propulsion of the 
pair and extract its Lighthill  energetic efficiency.
Further we demonstrate theoretically that it is possible to induce a velocity reversal by simply changing the magnetic properties of 
the particle or the rod geometric parameter. We confirm this effect by employing a longer nanorod in our self assembled microswimmer. We note that also our self-assembled pair does not need the presence of a boundary to propel, 
as the case of others magnetic propeller working at similar length scales. 

Our work can be further extended to other physical situations 
in order to explore how the efficiency or the directionality of the pair may be affected. For example all the results obtained until now have been based on the use of simple Newtonian fluids. However changing the medium with a viscoelastic fluid might lead to 
different developments where memory effects or time dependent phenomena become important and simple reciprocal motion may still lead to propulsion of the pair. 
Finally we stress that the realization of 
minimal, artificial microswimmers based on simple and externally tunable interactions, may be of great interest in different applications based on the controlled transport of drugs or chemicals into small pores or channels. The use of low frequency magnetic fields to actuate such functional micromachines has the advantages of not altering the dispersing medium (water) 
or affecting biological tissues. 
Finally, we provide a way to experimental characterize the energetic efficiency of such prototype, which may be of help in similar prototypes based on magnetic dipolar interactions. 

\section*{Conflicts of interest}
There are no conflicts to declare'.

\section*{Acknowledgements}
This work has received funding from the
Horizon 2020 research and innovation programme,
Grant Agreement No. 665440.
F.S. acknowledges support from MINECO under
project FIS2016-C2-1-P AEI/FEDER-EU.
I.P. acknowledges support from MINECO under
project FIS2015-67837-P and Generalitat de Catalunya under project 2017SGR-884 and SNF Project No.
200021-175719.
P.T. acknowledges the 
European Research Council (ENFORCE, No. 811234), MINECO 
(FIS2016-78507-C2-2-P, ERC2018-092827) 
and Generalitat de Catalunya under program "Icrea Academia".



\balance


\bibliography{biblio} 

\providecommand*{\mcitethebibliography}{\thebibliography}
\csname @ifundefined\endcsname{endmcitethebibliography}
{\let\endmcitethebibliography\endthebibliography}{}
\begin{mcitethebibliography}{61}
\providecommand*{\natexlab}[1]{#1}
\providecommand*{\mciteSetBstSublistMode}[1]{}
\providecommand*{\mciteSetBstMaxWidthForm}[2]{}
\providecommand*{\mciteBstWouldAddEndPuncttrue}
  {\def\EndOfBibitem{\unskip.}}
\providecommand*{\mciteBstWouldAddEndPunctfalse}
  {\let\EndOfBibitem\relax}
\providecommand*{\mciteSetBstMidEndSepPunct}[3]{}
\providecommand*{\mciteSetBstSublistLabelBeginEnd}[3]{}
\providecommand*{\EndOfBibitem}{}
\mciteSetBstSublistMode{f}
\mciteSetBstMaxWidthForm{subitem}
{(\emph{\alph{mcitesubitemcount}})}
\mciteSetBstSublistLabelBeginEnd{\mcitemaxwidthsubitemform\space}
{\relax}{\relax}

\bibitem[Bray(2000)]{Bra00}
D.~Bray, \emph{{Cell Movements}}, Garland Publishing, New York, 2000\relax
\mciteBstWouldAddEndPuncttrue
\mciteSetBstMidEndSepPunct{\mcitedefaultmidpunct}
{\mcitedefaultendpunct}{\mcitedefaultseppunct}\relax
\EndOfBibitem
\bibitem[Berg(2004)]{Ber04}
H.~C. Berg, \emph{{E. Coli in Motion}}, Springer-Verlag, New York, 2004\relax
\mciteBstWouldAddEndPuncttrue
\mciteSetBstMidEndSepPunct{\mcitedefaultmidpunct}
{\mcitedefaultendpunct}{\mcitedefaultseppunct}\relax
\EndOfBibitem
\bibitem[Happel and Brenner(1973)]{Happ}
J.~Happel and H.~Brenner, \emph{{Low Reynolds Number Hydrodynamics}},
  Noordhoff, Leiden, 1973\relax
\mciteBstWouldAddEndPuncttrue
\mciteSetBstMidEndSepPunct{\mcitedefaultmidpunct}
{\mcitedefaultendpunct}{\mcitedefaultseppunct}\relax
\EndOfBibitem
\bibitem[Purcell(1977)]{Pur97}
E.~M. Purcell, \emph{Am. J. Phys.}, 1977, \textbf{45}, 3\relax
\mciteBstWouldAddEndPuncttrue
\mciteSetBstMidEndSepPunct{\mcitedefaultmidpunct}
{\mcitedefaultendpunct}{\mcitedefaultseppunct}\relax
\EndOfBibitem
\bibitem[Lauga(2011)]{Lau11}
E.~Lauga, \emph{Soft Matter}, 2011, \textbf{7}, 3060–3065\relax
\mciteBstWouldAddEndPuncttrue
\mciteSetBstMidEndSepPunct{\mcitedefaultmidpunct}
{\mcitedefaultendpunct}{\mcitedefaultseppunct}\relax
\EndOfBibitem
\bibitem[Lauga and Powers(2009)]{Lau09}
E.~Lauga and T.~R. Powers, \emph{Rep. Prog. Phys.}, 2009, \textbf{72},
  096601\relax
\mciteBstWouldAddEndPuncttrue
\mciteSetBstMidEndSepPunct{\mcitedefaultmidpunct}
{\mcitedefaultendpunct}{\mcitedefaultseppunct}\relax
\EndOfBibitem
\bibitem[Tierno \emph{et~al.}(2008)Tierno, Golestanian, Pagonabarraga, and
  Sagués]{Tie082}
P.~Tierno, R.~Golestanian, I.~Pagonabarraga and F.~Sagués, \emph{The Journal
  of Physical Chemistry B}, 2008, \textbf{112}, 16525--16528\relax
\mciteBstWouldAddEndPuncttrue
\mciteSetBstMidEndSepPunct{\mcitedefaultmidpunct}
{\mcitedefaultendpunct}{\mcitedefaultseppunct}\relax
\EndOfBibitem
\bibitem[Sanchez \emph{et~al.}(2011)Sanchez, Solovev, Harazim, and
  Schmidt]{San11}
S.~Sanchez, A.~A. Solovev, S.~M. Harazim and O.~G. Schmidt, \emph{Journal of
  the American Chemical Society}, 2011, \textbf{133}, 701--703\relax
\mciteBstWouldAddEndPuncttrue
\mciteSetBstMidEndSepPunct{\mcitedefaultmidpunct}
{\mcitedefaultendpunct}{\mcitedefaultseppunct}\relax
\EndOfBibitem
\bibitem[Peyer \emph{et~al.}(2013)Peyer, Zhang, and Nelson]{Pey13}
K.~E. Peyer, L.~Zhang and B.~J. Nelson, \emph{Nanoscale}, 2013, \textbf{5},
  1259\relax
\mciteBstWouldAddEndPuncttrue
\mciteSetBstMidEndSepPunct{\mcitedefaultmidpunct}
{\mcitedefaultendpunct}{\mcitedefaultseppunct}\relax
\EndOfBibitem
\bibitem[Kim \emph{et~al.}(2013)Kim, Qiu, Kim, Ghanbari, Moon, Zhang, Nelson,
  and Choi]{Kim13}
S.~Kim, F.~Qiu, S.~Kim, A.~Ghanbari, C.~Moon, L.~Zhang, B.~J. Nelson and
  H.~Choi, \emph{Adv. Mater.}, 2013, \textbf{25}, 5863\relax
\mciteBstWouldAddEndPuncttrue
\mciteSetBstMidEndSepPunct{\mcitedefaultmidpunct}
{\mcitedefaultendpunct}{\mcitedefaultseppunct}\relax
\EndOfBibitem
\bibitem[Barbot \emph{et~al.}(2016)Barbot, Decanini, and Hwang]{Bar16}
A.~Barbot, D.~Decanini and G.~Hwang, \emph{Scientific Reports}, 2016,
  \textbf{6}, 19041\relax
\mciteBstWouldAddEndPuncttrue
\mciteSetBstMidEndSepPunct{\mcitedefaultmidpunct}
{\mcitedefaultendpunct}{\mcitedefaultseppunct}\relax
\EndOfBibitem
\bibitem[Bechinger \emph{et~al.}(2016)Bechinger, Di~Leonardo, L\"owen,
  Reichhardt, Volpe, and Volpe]{Bec16}
C.~Bechinger, R.~Di~Leonardo, H.~L\"owen, C.~Reichhardt, G.~Volpe and G.~Volpe,
  \emph{Rev. Mod. Phys.}, 2016, \textbf{88}, 045006\relax
\mciteBstWouldAddEndPuncttrue
\mciteSetBstMidEndSepPunct{\mcitedefaultmidpunct}
{\mcitedefaultendpunct}{\mcitedefaultseppunct}\relax
\EndOfBibitem
\bibitem[Ma \emph{et~al.}(2015)Ma, Yang, Zhao, and Wu]{FMa15}
F.~Ma, X.~Yang, H.~Zhao and N.~Wu, \emph{Phys. Rev. Lett.}, 2015, \textbf{115},
  208302\relax
\mciteBstWouldAddEndPuncttrue
\mciteSetBstMidEndSepPunct{\mcitedefaultmidpunct}
{\mcitedefaultendpunct}{\mcitedefaultseppunct}\relax
\EndOfBibitem
\bibitem[Ma \emph{et~al.}(2015)Ma, Wang, Wu, and Wu]{FMa152}
F.~Ma, S.~Wang, D.~T. Wu and N.~Wu, \emph{Proc. Natl. Acad. Sci. USA}, 2015,
  \textbf{112}, 6307\relax
\mciteBstWouldAddEndPuncttrue
\mciteSetBstMidEndSepPunct{\mcitedefaultmidpunct}
{\mcitedefaultendpunct}{\mcitedefaultseppunct}\relax
\EndOfBibitem
\bibitem[Dreyfus \emph{et~al.}(2005)Dreyfus, Baudry, Roper, Fermigier, Stone,
  and Bibette]{Dre05}
R.~Dreyfus, J.~Baudry, M.~L. Roper, M.~Fermigier, H.~A. Stone and J.~Bibette,
  \emph{Nat. Comm.}, 2005, \textbf{437}, 862\relax
\mciteBstWouldAddEndPuncttrue
\mciteSetBstMidEndSepPunct{\mcitedefaultmidpunct}
{\mcitedefaultendpunct}{\mcitedefaultseppunct}\relax
\EndOfBibitem
\bibitem[Zhang \emph{et~al.}(2009)Zhang, Abbott, Dong, Peyer, Kratochvil,
  Zhang, Bergeles, and Nelson]{Zha09}
L.~Zhang, J.~J. Abbott, L.~Dong, K.~E. Peyer, B.~E. Kratochvil, H.~Zhang,
  C.~Bergeles and B.~J. Nelson, \emph{Nano Lett.}, 2009, \textbf{9}, 3663\relax
\mciteBstWouldAddEndPuncttrue
\mciteSetBstMidEndSepPunct{\mcitedefaultmidpunct}
{\mcitedefaultendpunct}{\mcitedefaultseppunct}\relax
\EndOfBibitem
\bibitem[Ghosh and Fischer(2009)]{Gho09}
A.~Ghosh and P.~Fischer, \emph{Nano Letters}, 2009, \textbf{9},
  2243--2245\relax
\mciteBstWouldAddEndPuncttrue
\mciteSetBstMidEndSepPunct{\mcitedefaultmidpunct}
{\mcitedefaultendpunct}{\mcitedefaultseppunct}\relax
\EndOfBibitem
\bibitem[Snezhko \emph{et~al.}(2009)Snezhko, Belkin, Aranson, and Kwok]{Sne09}
A.~Snezhko, M.~Belkin, I.~S. Aranson and W.-K. Kwok, \emph{Phys. Rev. Lett.},
  2009, \textbf{102}, 118103\relax
\mciteBstWouldAddEndPuncttrue
\mciteSetBstMidEndSepPunct{\mcitedefaultmidpunct}
{\mcitedefaultendpunct}{\mcitedefaultseppunct}\relax
\EndOfBibitem
\bibitem[Volpe \emph{et~al.}()Volpe, Buttinoni, Vogt, Kummerer, and
  Bechinger]{Vol11}
G.~Volpe, I.~Buttinoni, D.~Vogt, H.-J. Kummerer and C.~Bechinger, \emph{Soft
  Matter}, \textbf{7}, 8810\relax
\mciteBstWouldAddEndPuncttrue
\mciteSetBstMidEndSepPunct{\mcitedefaultmidpunct}
{\mcitedefaultendpunct}{\mcitedefaultseppunct}\relax
\EndOfBibitem
\bibitem[Palagi \emph{et~al.}(2016)Palagi, Mark, Reigh, Melde, Qiu, Zeng,
  Parmeggiani, Martella, Sanchez-Castillo, Kapernaum, Giesselmann, Wiersma,
  Lauga, and Fischer]{Pal16}
S.~Palagi, A.~G. Mark, S.~Y. Reigh, K.~Melde, T.~Qiu, H.~Zeng, C.~Parmeggiani,
  D.~Martella, A.~Sanchez-Castillo, N.~Kapernaum, F.~Giesselmann, D.~S.
  Wiersma, E.~Lauga and P.~Fischer, \emph{Nat. Mater.}, 2016, \textbf{15},
  647\relax
\mciteBstWouldAddEndPuncttrue
\mciteSetBstMidEndSepPunct{\mcitedefaultmidpunct}
{\mcitedefaultendpunct}{\mcitedefaultseppunct}\relax
\EndOfBibitem
\bibitem[W.~Wang(2012)]{Wan12}
M.~H. T.~M. W.~Wang, L.A.~Castro, \emph{ACS Nano}, 2012, \textbf{72},
  6122--6132\relax
\mciteBstWouldAddEndPuncttrue
\mciteSetBstMidEndSepPunct{\mcitedefaultmidpunct}
{\mcitedefaultendpunct}{\mcitedefaultseppunct}\relax
\EndOfBibitem
\bibitem[Xu \emph{et~al.}(2014)Xu, Soto, Gao, Garcia-Gradilla, Li, Zhang, and
  Wang]{XuT14}
T.~Xu, F.~Soto, W.~Gao, V.~Garcia-Gradilla, J.~Li, X.~Zhang and J.~Wang,
  \emph{J. Am. Chem. Soc.}, 2014, \textbf{136}, 8552--8555\relax
\mciteBstWouldAddEndPuncttrue
\mciteSetBstMidEndSepPunct{\mcitedefaultmidpunct}
{\mcitedefaultendpunct}{\mcitedefaultseppunct}\relax
\EndOfBibitem
\bibitem[Tailin~Xu(2017)]{Tai17}
X.~Z. Tailin~Xu, Li-Ping~Xu, \emph{Applied Materials Today}, 2017, \textbf{9},
  493--503\relax
\mciteBstWouldAddEndPuncttrue
\mciteSetBstMidEndSepPunct{\mcitedefaultmidpunct}
{\mcitedefaultendpunct}{\mcitedefaultseppunct}\relax
\EndOfBibitem
\bibitem[Paxton \emph{et~al.}(2004)Paxton, Kistler, Olmeda, Sen, Angelo, Cao,
  Mallouk, Lammert, and Crespi]{Pax04}
W.~F. Paxton, K.~C. Kistler, C.~C. Olmeda, A.~Sen, S.~K.~S. Angelo, Y.~Cao,
  T.~E. Mallouk, P.~E. Lammert and V.~H. Crespi, \emph{J. Am. Chem. Soc.},
  2004, \textbf{126}, 13424\relax
\mciteBstWouldAddEndPuncttrue
\mciteSetBstMidEndSepPunct{\mcitedefaultmidpunct}
{\mcitedefaultendpunct}{\mcitedefaultseppunct}\relax
\EndOfBibitem
\bibitem[Howse \emph{et~al.}(2007)Howse, Jones, Ryan, Gough, Vafabakhsh, and
  Golestanian]{How07}
J.~R. Howse, R.~A.~L. Jones, A.~J. Ryan, T.~Gough, R.~Vafabakhsh and
  R.~Golestanian, \emph{Phys. Rev. Lett.}, 2007, \textbf{99}, 048102\relax
\mciteBstWouldAddEndPuncttrue
\mciteSetBstMidEndSepPunct{\mcitedefaultmidpunct}
{\mcitedefaultendpunct}{\mcitedefaultseppunct}\relax
\EndOfBibitem
\bibitem[Uspal \emph{et~al.}(2016)Uspal, Popescu, Dietrich, and
  Tasinkevych]{Usp16}
W.~E. Uspal, M.~N. Popescu, S.~Dietrich and M.~Tasinkevych, \emph{Phys. Rev.
  Lett.}, 2016, \textbf{117}, 048002\relax
\mciteBstWouldAddEndPuncttrue
\mciteSetBstMidEndSepPunct{\mcitedefaultmidpunct}
{\mcitedefaultendpunct}{\mcitedefaultseppunct}\relax
\EndOfBibitem
\bibitem[Gao \emph{et~al.}(2011)Gao, Sattayasamitsathit, Orozco, and
  Wang]{Gao2011}
W.~Gao, S.~Sattayasamitsathit, J.~Orozco and J.~Wang, \emph{J. Am. Chem. Soc.},
  2011, \textbf{133}, 11862\relax
\mciteBstWouldAddEndPuncttrue
\mciteSetBstMidEndSepPunct{\mcitedefaultmidpunct}
{\mcitedefaultendpunct}{\mcitedefaultseppunct}\relax
\EndOfBibitem
\bibitem[Palacci \emph{et~al.}(2013)Palacci, Sacanna, Vatchinsky, Chaikin, and
  Pine]{Palacci2013}
J.~Palacci, S.~Sacanna, A.~Vatchinsky, P.~M. Chaikin and D.~J. Pine, \emph{J.
  Am. Chem. Soc.}, 2013, \textbf{135}, 15978\relax
\mciteBstWouldAddEndPuncttrue
\mciteSetBstMidEndSepPunct{\mcitedefaultmidpunct}
{\mcitedefaultendpunct}{\mcitedefaultseppunct}\relax
\EndOfBibitem
\bibitem[Massana-Cid \emph{et~al.}(2018)Massana-Cid, Codina, Pagonabarraga, and
  Tierno]{Massana2018}
H.~Massana-Cid, J.~Codina, I.~Pagonabarraga and P.~Tierno, \emph{Proc. Natl.
  Acad. Sci. U. S. A.}, 2018, \textbf{115}, 10618\relax
\mciteBstWouldAddEndPuncttrue
\mciteSetBstMidEndSepPunct{\mcitedefaultmidpunct}
{\mcitedefaultendpunct}{\mcitedefaultseppunct}\relax
\EndOfBibitem
\bibitem[Tierno \emph{et~al.}(2008)Tierno, Golestanian, Pagonabarraga, and
  Sagu\'es]{Tie08}
P.~Tierno, R.~Golestanian, I.~Pagonabarraga and F.~Sagu\'es, \emph{Phys. Rev.
  Lett.}, 2008, \textbf{101}, 218304\relax
\mciteBstWouldAddEndPuncttrue
\mciteSetBstMidEndSepPunct{\mcitedefaultmidpunct}
{\mcitedefaultendpunct}{\mcitedefaultseppunct}\relax
\EndOfBibitem
\bibitem[Cheang and Kima(2016)]{Cheang2016}
U.~K. Cheang and M.~J. Kima, \emph{Appl. Phys. Lett.}, 2016, \textbf{109},
  034101\relax
\mciteBstWouldAddEndPuncttrue
\mciteSetBstMidEndSepPunct{\mcitedefaultmidpunct}
{\mcitedefaultendpunct}{\mcitedefaultseppunct}\relax
\EndOfBibitem
\bibitem[Cheang \emph{et~al.}(2014)Cheang, Meshkati, Kim, Kim, and
  Fu]{Cheang2014}
U.~K. Cheang, F.~Meshkati, D.~Kim, M.~J. Kim and H.~C. Fu, \emph{Phys. Rev. E},
  2014, \textbf{90}, 033007\relax
\mciteBstWouldAddEndPuncttrue
\mciteSetBstMidEndSepPunct{\mcitedefaultmidpunct}
{\mcitedefaultendpunct}{\mcitedefaultseppunct}\relax
\EndOfBibitem
\bibitem[Sachs \emph{et~al.}(2018)Sachs, Morozov, Kenneth, Qiu, Segreto,
  Fischer, and Leshansky]{Sachs2018}
J.~Sachs, K.~I. Morozov, O.~Kenneth, T.~Qiu, N.~Segreto, P.~Fischer and A.~M.
  Leshansky, \emph{Phys. Rev. E}, 2018, \textbf{98}, 063105\relax
\mciteBstWouldAddEndPuncttrue
\mciteSetBstMidEndSepPunct{\mcitedefaultmidpunct}
{\mcitedefaultendpunct}{\mcitedefaultseppunct}\relax
\EndOfBibitem
\bibitem[Tottori and Nelson(2018)]{Tottori2018}
S.~Tottori and B.~J. Nelson, \emph{Small}, 2018, \textbf{14}, 1800722\relax
\mciteBstWouldAddEndPuncttrue
\mciteSetBstMidEndSepPunct{\mcitedefaultmidpunct}
{\mcitedefaultendpunct}{\mcitedefaultseppunct}\relax
\EndOfBibitem
\bibitem[Mirzae \emph{et~al.}(2018)Mirzae, Dubrovski, Kenneth, Morozov, and
  Leshansky]{Mirza2018}
Y.~Mirzae, O.~Dubrovski, O.~Kenneth, K.~I. Morozov and A.~M. Leshansky,
  \emph{Science Robotics}, 2018, \textbf{3}, eaas8713\relax
\mciteBstWouldAddEndPuncttrue
\mciteSetBstMidEndSepPunct{\mcitedefaultmidpunct}
{\mcitedefaultendpunct}{\mcitedefaultseppunct}\relax
\EndOfBibitem
\bibitem[Cohen \emph{et~al.}(2019)Cohen, Rubinstein, Kenneth, and
  Leshansky]{Cohen2019}
K.-J. Cohen, B.~Y. Rubinstein, O.~Kenneth and A.~M. Leshansky, \emph{Phys. Rev.
  Applied}, 2019, \textbf{12}, 014025\relax
\mciteBstWouldAddEndPuncttrue
\mciteSetBstMidEndSepPunct{\mcitedefaultmidpunct}
{\mcitedefaultendpunct}{\mcitedefaultseppunct}\relax
\EndOfBibitem
\bibitem[Calero \emph{et~al.}(2019)Calero, Garc\'ia-Torres, Ortiz-Ambriz,
  Sagu\'es, Pagonabarraga, and Tierno]{Calero2019}
C.~Calero, J.~Garc\'ia-Torres, A.~Ortiz-Ambriz, F.~Sagu\'es, I.~Pagonabarraga
  and P.~Tierno, \emph{Nanoscale}, 2019, \textbf{11}, 18723\relax
\mciteBstWouldAddEndPuncttrue
\mciteSetBstMidEndSepPunct{\mcitedefaultmidpunct}
{\mcitedefaultendpunct}{\mcitedefaultseppunct}\relax
\EndOfBibitem
\bibitem[Martinez-Pedrero \emph{et~al.}(2016)Martinez-Pedrero, Cebers, and
  Tierno]{Mar16}
F.~Martinez-Pedrero, A.~Cebers and P.~Tierno, \emph{Phys. Rev. Applied}, 2016,
  \textbf{6}, 034002\relax
\mciteBstWouldAddEndPuncttrue
\mciteSetBstMidEndSepPunct{\mcitedefaultmidpunct}
{\mcitedefaultendpunct}{\mcitedefaultseppunct}\relax
\EndOfBibitem
\bibitem[Martinez-Pedrero \emph{et~al.}(2016)Martinez-Pedrero, Cebers, and
  Tierno]{Mart16}
F.~Martinez-Pedrero, A.~Cebers and P.~Tierno, \emph{Soft Matter}, 2016,
  \textbf{12}, 3688--3695\relax
\mciteBstWouldAddEndPuncttrue
\mciteSetBstMidEndSepPunct{\mcitedefaultmidpunct}
{\mcitedefaultendpunct}{\mcitedefaultseppunct}\relax
\EndOfBibitem
\bibitem[Garc\'ia-Torres \emph{et~al.}(2018)Garc\'ia-Torres, Calero, Sagu\'es,
  Pagonabarraga, and Tierno]{Jos18}
J.~Garc\'ia-Torres, C.~Calero, F.~Sagu\'es, I.~Pagonabarraga and P.~Tierno,
  \emph{Nat. Comm.}, 2018, \textbf{9}, 1663\relax
\mciteBstWouldAddEndPuncttrue
\mciteSetBstMidEndSepPunct{\mcitedefaultmidpunct}
{\mcitedefaultendpunct}{\mcitedefaultseppunct}\relax
\EndOfBibitem
\bibitem[Tierno \emph{et~al.}(2009)Tierno, Schreiber, Zimmermann, and
  Fischer]{Tie091}
P.~Tierno, S.~Schreiber, W.~Zimmermann and T.~M. Fischer, \emph{Journal of the
  American Chemical Society}, 2009, \textbf{131}, 5366--5367\relax
\mciteBstWouldAddEndPuncttrue
\mciteSetBstMidEndSepPunct{\mcitedefaultmidpunct}
{\mcitedefaultendpunct}{\mcitedefaultseppunct}\relax
\EndOfBibitem
\bibitem[Morimoto \emph{et~al.}(2008)Morimoto, Ukai, Nagaoka, Grobert, and
  Maekawa]{Mor08}
H.~Morimoto, T.~Ukai, Y.~Nagaoka, N.~Grobert and T.~Maekawa, \emph{Phys. Rev.
  E}, 2008, \textbf{78}, 021403\relax
\mciteBstWouldAddEndPuncttrue
\mciteSetBstMidEndSepPunct{\mcitedefaultmidpunct}
{\mcitedefaultendpunct}{\mcitedefaultseppunct}\relax
\EndOfBibitem
\bibitem[Zhang \emph{et~al.}(2010)Zhang, Petit, Lu, Kratochvil, Peyer, Pei,
  Lou, and Nelson]{Zha10}
L.~Zhang, T.~Petit, Y.~Lu, B.~E. Kratochvil, K.~E. Peyer, R.~Pei, J.~Lou and
  B.~J. Nelson, \emph{ACS Nano}, 2010, \textbf{4}, 6228--6234\relax
\mciteBstWouldAddEndPuncttrue
\mciteSetBstMidEndSepPunct{\mcitedefaultmidpunct}
{\mcitedefaultendpunct}{\mcitedefaultseppunct}\relax
\EndOfBibitem
\bibitem[Schmid \emph{et~al.}(2010)Schmid, Schneider, Franke, and
  Alexander-Katz]{Sin10}
C.~E. S.~L. Schmid, M.~F. Schneider, T.~Franke and A.~Alexander-Katz,
  \emph{Proc. Natl. Acad. Sci. U.S.A.}, 2010, \textbf{107}, 535\relax
\mciteBstWouldAddEndPuncttrue
\mciteSetBstMidEndSepPunct{\mcitedefaultmidpunct}
{\mcitedefaultendpunct}{\mcitedefaultseppunct}\relax
\EndOfBibitem
\bibitem[Garcia-Torres \emph{et~al.}(2017)Garcia-Torres, Serra, Tierno, Alcobe,
  and Valles]{Jos17}
J.~Garcia-Torres, A.~Serra, P.~Tierno, X.~Alcobe and E.~Valles, \emph{ACS Appl
  Mater Interfaces}, 2017, \textbf{9}, 23859\relax
\mciteBstWouldAddEndPuncttrue
\mciteSetBstMidEndSepPunct{\mcitedefaultmidpunct}
{\mcitedefaultendpunct}{\mcitedefaultseppunct}\relax
\EndOfBibitem
\bibitem[Bricard \emph{et~al.}(2013)Bricard, Caussin, Desreumaux, Dauchot, and
  Bartolo]{Bri13}
A.~Bricard, J.-B. Caussin, N.~Desreumaux, O.~Dauchot and D.~Bartolo,
  \emph{Nature}, 2013, \textbf{503}, 95\relax
\mciteBstWouldAddEndPuncttrue
\mciteSetBstMidEndSepPunct{\mcitedefaultmidpunct}
{\mcitedefaultendpunct}{\mcitedefaultseppunct}\relax
\EndOfBibitem
\bibitem[Martinez-Pedrero \emph{et~al.}(2015)Martinez-Pedrero, Ortiz-Ambriz,
  Pagonabarraga, and Tierno]{Mar15}
F.~Martinez-Pedrero, A.~Ortiz-Ambriz, I.~Pagonabarraga and P.~Tierno,
  \emph{Phys. Rev. Lett.}, 2015, \textbf{115}, 138301\relax
\mciteBstWouldAddEndPuncttrue
\mciteSetBstMidEndSepPunct{\mcitedefaultmidpunct}
{\mcitedefaultendpunct}{\mcitedefaultseppunct}\relax
\EndOfBibitem
\bibitem[Delmotte \emph{et~al.}(2017)Delmotte, Driscoll, Chaikin, and
  Donev]{Del17}
B.~Delmotte, M.~Driscoll, P.~Chaikin and A.~Donev, \emph{Phys. Rev. Fluids},
  2017, \textbf{2}, 092301\relax
\mciteBstWouldAddEndPuncttrue
\mciteSetBstMidEndSepPunct{\mcitedefaultmidpunct}
{\mcitedefaultendpunct}{\mcitedefaultseppunct}\relax
\EndOfBibitem
\bibitem[Weeks and Chandler(1971)]{WCA}
J.~D. Weeks and D.~Chandler, \emph{J. Chem. Phys.}, 1971, \textbf{54},
  5237\relax
\mciteBstWouldAddEndPuncttrue
\mciteSetBstMidEndSepPunct{\mcitedefaultmidpunct}
{\mcitedefaultendpunct}{\mcitedefaultseppunct}\relax
\EndOfBibitem
\bibitem[Bailey \emph{et~al.}(2009)Bailey, Lowe, Pagonabarraga, and
  Lagomarsino]{Bai09}
A.~G. Bailey, C.~P. Lowe, I.~Pagonabarraga and M.~C. Lagomarsino, \emph{Phys.
  Rev. E}, 2009, \textbf{80}, 046707\relax
\mciteBstWouldAddEndPuncttrue
\mciteSetBstMidEndSepPunct{\mcitedefaultmidpunct}
{\mcitedefaultendpunct}{\mcitedefaultseppunct}\relax
\EndOfBibitem
\bibitem[Gutman and Or(2016)]{Emi16}
E.~Gutman and Y.~Or, \emph{Phys. Rev. E}, 2016, \textbf{93}, 063105\relax
\mciteBstWouldAddEndPuncttrue
\mciteSetBstMidEndSepPunct{\mcitedefaultmidpunct}
{\mcitedefaultendpunct}{\mcitedefaultseppunct}\relax
\EndOfBibitem
\bibitem[Lig(1975)]{Lig75}
Mathematical Biofluiddynamics, 1975\relax
\mciteBstWouldAddEndPuncttrue
\mciteSetBstMidEndSepPunct{\mcitedefaultmidpunct}
{\mcitedefaultendpunct}{\mcitedefaultseppunct}\relax
\EndOfBibitem
\bibitem[Purcell(1997)]{Pu97}
E.~M. Purcell, \emph{Proc. Natl Acad. Sci. USA}, 1997, \textbf{94},
  11307–11311\relax
\mciteBstWouldAddEndPuncttrue
\mciteSetBstMidEndSepPunct{\mcitedefaultmidpunct}
{\mcitedefaultendpunct}{\mcitedefaultseppunct}\relax
\EndOfBibitem
\bibitem[Shapere and Wilczek(1989)]{Sha89}
A.~Shapere and F.~Wilczek, \emph{J. Fluid Mech.}, 1989, \textbf{198}, 587\relax
\mciteBstWouldAddEndPuncttrue
\mciteSetBstMidEndSepPunct{\mcitedefaultmidpunct}
{\mcitedefaultendpunct}{\mcitedefaultseppunct}\relax
\EndOfBibitem
\bibitem[Ostermana and Vilfan(2011)]{Ost11}
N.~Ostermana and A.~Vilfan, \emph{Proc. Natl Acad. Sci. USA}, 2011,
  \textbf{108}, 15727\relax
\mciteBstWouldAddEndPuncttrue
\mciteSetBstMidEndSepPunct{\mcitedefaultmidpunct}
{\mcitedefaultendpunct}{\mcitedefaultseppunct}\relax
\EndOfBibitem
\bibitem[Becker \emph{et~al.}(2003)Becker, Koehler, and Stone]{Bec03}
L.~E. Becker, S.~A. Koehler and H.~A. Stone, \emph{J. Fluid Mech.}, 2003,
  \textbf{490}, 15\relax
\mciteBstWouldAddEndPuncttrue
\mciteSetBstMidEndSepPunct{\mcitedefaultmidpunct}
{\mcitedefaultendpunct}{\mcitedefaultseppunct}\relax
\EndOfBibitem
\bibitem[Passov and Ora(2012)]{Pas12}
E.~Passov and Y.~Ora, \emph{Eur. Phys. J. E}, 2012, \textbf{35}, 78\relax
\mciteBstWouldAddEndPuncttrue
\mciteSetBstMidEndSepPunct{\mcitedefaultmidpunct}
{\mcitedefaultendpunct}{\mcitedefaultseppunct}\relax
\EndOfBibitem
\bibitem[Wiezel and Or(2016)]{Wie16}
O.~Wiezel and Y.~Or, \emph{Proc. R. Soc. A}, 2016, \textbf{472}, 20160425\relax
\mciteBstWouldAddEndPuncttrue
\mciteSetBstMidEndSepPunct{\mcitedefaultmidpunct}
{\mcitedefaultendpunct}{\mcitedefaultseppunct}\relax
\EndOfBibitem
\bibitem[Golestanian and Ajdari(2008)]{Ram08}
R.~Golestanian and A.~Ajdari, \emph{Phys. Rev. E}, 2008, \textbf{77},
  036308\relax
\mciteBstWouldAddEndPuncttrue
\mciteSetBstMidEndSepPunct{\mcitedefaultmidpunct}
{\mcitedefaultendpunct}{\mcitedefaultseppunct}\relax
\EndOfBibitem
\bibitem[Ishimoto and Gaffney(2014)]{Ishi14}
K.~Ishimoto and E.~A. Gaffney, \emph{Phys. Rev. E}, 2014, \textbf{90},
  012704\relax
\mciteBstWouldAddEndPuncttrue
\mciteSetBstMidEndSepPunct{\mcitedefaultmidpunct}
{\mcitedefaultendpunct}{\mcitedefaultseppunct}\relax
\EndOfBibitem
\bibitem[Raz and Leshansky(2008)]{Raz08}
O.~Raz and A.~M. Leshansky, \emph{Phys. Rev. E}, 2008, \textbf{77},
  055305\relax
\mciteBstWouldAddEndPuncttrue
\mciteSetBstMidEndSepPunct{\mcitedefaultmidpunct}
{\mcitedefaultendpunct}{\mcitedefaultseppunct}\relax
\EndOfBibitem
\end{mcitethebibliography}
\bibliographystyle{rsc} 

\end{document}